\documentclass[aps,pre,showpacs,twocolumn,superscriptaddress,letterpaper]{revtex4}
\usepackage{mathrsfs}
\usepackage{amsmath}
\usepackage{bm}
\usepackage{color}
\usepackage{graphicx}
\usepackage{hyperref}
\hypersetup{pdfpagemode=UseNone,pdfstartview=FitH,pdfstartpage=1}

\begin{document}


\title{Local Gyrokinetic Study of Electrostatic Microinstabilities in Dipole Plasmas}
\author{Hua-sheng Xie}
\email[]{Email: huashengxie@gmail.com} \affiliation{Fusion
Simulation Center, State Key Laboratory of Nuclear Physics and
Technology, School of Physics, Peking University, Beijing 100871,
China}
\author{Yi Zhang}
\affiliation{Fusion Simulation Center, State Key Laboratory of
Nuclear Physics and Technology, School of Physics, Peking
University, Beijing 100871, China}
\author{Zi-cong Huang}
\affiliation{Fusion Simulation Center, State Key Laboratory of
Nuclear Physics and Technology, School of Physics, Peking
University, Beijing 100871, China}
\author{Wei-ke Ou}
\affiliation{Fusion Simulation Center, State Key Laboratory of
Nuclear Physics and Technology, School of Physics, Peking
University, Beijing 100871, China}
\author{Bo Li}\email[Corresponding author.~]{Email: bli@pku.edu.cn}\affiliation{Fusion
Simulation Center, State Key Laboratory of Nuclear Physics and
Technology, School of Physics, Peking University, Beijing 100871,
China}

\date{\today}

\begin{abstract}
A linear gyrokinetic particle-in-cell scheme, which is valid for
arbitrary perpendicular wavelength $k_\perp\rho_i$ and includes the
parallel dynamic along the field line, is developed to study the
local electrostatic drift modes in point and ring dipole plasmas. We
find the most unstable mode in this system can be either electron
mode or ion mode. The properties and relations of these modes are
studied in detail as a function of $k_\perp\rho_i$, the density
gradient $\kappa_n$, the temperature gradient $\kappa_T$, electron
to ion temperature ratio $\tau=T_e/T_i$, and mass ratio $m_i/m_e$.
For conventional weak gradient parameters, the mode is on ground
state (with eigenstate number $l=0$) and especially
$k_\parallel\sim0$ for small $k_\perp\rho_i$. Thus, bounce averaged
dispersion relation is also derived for comparison. For strong
gradient and large $k_\perp\rho_i$, most interestingly, higher order
eigenstate modes with even (e.g., $l=2,4$) or odd (e.g., $l=1$)
parity can be most unstable, which is not expected by previous
studies. High order eigenstate can also easily be most unstable at
weak gradient when $\tau>10$. This work can be particularly
important to understand the turbulent transport in laboratory and
space magnetosphere.
\end{abstract}


\maketitle

\section{Introduction}\label{sec:intro}
Dipole magnetic fields widely exist in the Universe, such as in the
planetary magnetospheres. The idea of using strong dipole field
configuration for magnetic confinement of laboratory plasmas for
fusion is proposed theoretically by Hasegawa
\cite{Hasegawa1987,Hasegawa1990} and several experimental devices
have also been built since then, such as the Levitated Dipole
Experiment (LDX) \cite{Boxer2008,Boxer2010,Garnier2017} at MIT, the
Collisionless Terrella Experiment (CTX) \cite{Levitt2002} at
Columbia University and Ring Trap-1 (RT-1)
\cite{Yoshida2006,Yoshida2013} at the University of Tokyo. The
dipole configuration is also used to confine electron-positron pair
plasmas in the laboratory\cite{Sarri2015}. Typical charged particle
trajectories under ideal dipole field have good confinement
features. The collision and electrostatic or electromagnetic
turbulence can break this ideal confinement. Previous theoretical
and experimental results in
Refs.\cite{Boxer2008,Boxer2010,Garnier2017} show that the dipole
confinement is good even under the stochastic motion of particles
due to the collision and turbulence, at least under the present
laboratory low temperature and density parameters. However, the
confinement properties of dipole field under fusion parameters,
e.g., high temperature and density and the gradient of them, are
still open questions.

In this work, we develop a local gyrokinetic particle code to
understand the linear behaviors of electrostatic microinstabilities
in point (ideal) and ring dipole plasmas. In contrast to previous
studies\cite{Kesner1998,Kesner2000,Kesner2002,Simakov2001,Helander2016},
we do not limit our study to small $k_\perp\rho_i$, where $k_\perp$
is the perpendicular wavevector and $\rho_i$ is the ion Lamor
radius. Comprehensive investigations of the linear features of the
electrostatic drift modes can be important to understand the
nonlinear physics\cite{Horton1999}, such as the simulation results
of the typical electrostatic turbulent transport features in ring
dipole configuration in Refs.\cite{Kobayashi2009,Kobayashi2010}.
Several electromagnetic studies
\cite{Simakov2002,Dettrick2003,Porazik2011,Mager2017} of the
Alfv\'enic drift modes in dipole configuration also exist, which is
not the aim of the present work. We also noticed that
Ref.\cite{Zhao2001} had attempted to build a particle-in-cell model
similar to our present work using gyrokinetic ion but
bounce-averaged electron model in local point dipole configuration.

In the following sections, we provide a comprehensive derivations of
our model. Sec.\ref{sec:model} gives the linearized gyrokinetic
model equations. Sec.\ref{sec:dip_op} provides the details of the
dipole coordinate system and operators. Sec.\ref{sec:dr} gives the
relevant zero-dimensional dispersion relation. Sec.\ref{sec:pic}
discusses some details of our particle-in-cell model.
Sec.\ref{sec:bech} benchmarks our simulation model.
Sec.\ref{sec:sim_result} shows the details of our simulation
results. Sec.\ref{sec:summ} summarizes the present study.

\section{Linear Gyrokinetic Model}\label{sec:model}

We use standard linear gyrokinetic
model\cite{Antonsen1980,Chen1991}, which can describe the low
frequency physics accurately under the assumptions:
$\omega/\Omega_{ci}\sim\rho_i/L\sim k_\parallel/k_\perp\ll1$.
Assuming Maxwellian equilibrium distribution function
$F_{0}=n_0F_M$, with $F_M=(\frac{m}{2\pi T})^{3/2}e^{-m\epsilon/T}$,
$\epsilon=v^2/2$, $\mu=v_\perp^2/2B$, the perturbation distribution
function after gyrophase average is
\begin{equation}\label{eq:gk_f0}
  f_{s} = \frac{q_s}{m_s} \frac{\partial F^s_{0}}{\partial\epsilon} \phi + J_{0}(k_\perp\rho_{s})
  h_{s}.
\end{equation}
The non-adiabatic gyrokinetic response $h_s$ satisfies
\begin{equation}\label{eq:gk_dh}
(\omega - \omega_{Ds} + i v_{\parallel} {\bf b} \cdot {\bf \nabla})
h_{s}
  = - (\omega - \omega_{*s}^T)
  \frac{\partial F^s_{0}}{\partial\epsilon} \frac{q_s}{m_s}J_{0} \phi,
\end{equation}
where the first kind Bessel function $J_0$ comes from  gyrophase
average, and the parameters are
\begin{eqnarray*}
  \rho_{s} &=& \frac{v_{\perp}}{\Omega_{s}}, ~~~~\Omega_s=\frac{q_s B}{m_sc}, \\
  \frac{\partial F^s_{0}}{\partial\epsilon} &=& -\frac{m_sF_0^s}{T_s}, ~~~{\bf b} = {\bf B} / B,\\
  \omega_{*s}^T &=& \frac{{\bf k}_{\perp} \times {\bf b}
\cdot {\bf \nabla} F^{s}_{0} }{ -\Omega_{s}F^{s}_{0\epsilon}}, \\
  \omega_{Ds} &=& {\bf k}_{\perp} \cdot {\bf v}_d={\bf k}_{\perp} \cdot {\bf b} \times \frac{\mu \nabla B
+ v_{\parallel}^{2} {\bf b} \cdot \nabla {\bf b}}{\Omega_{s}}.
\end{eqnarray*}
Here, $s=i,e$ represents particle species, and the collision term is
neglected. ${\bf B}$ is the magnetic field, and $q_s$, $m_s$, $T_s$,
$\Omega_s$, $\rho_s$, $\omega_{*s}^T$ and $\omega_{Ds}$ are the
charge, mass, temperature, cyclotron frequency, gyroradius,
diamagnetic drift frequency and magnetic (gradient and curvature)
drift frequency for the species $s$, respectively.

In electrostatic case, the gyrokinetic system is closed by
quasi-neutrality condition (Poisson equation)
\begin{equation}\label{eq:gk_n}
  \sum_\alpha q_\alpha\int f_{\alpha} d^{3} v =0,
\end{equation}
where the notation for velocity integral $\int d^3 v\equiv
2\pi\int\frac{B}{|v_\parallel|}d\epsilon d\lambda$, with pitch angle
variable $\lambda\equiv\mu B_0/\epsilon$. In Eq.(\ref{eq:gk_n}), we
have assumed the Debye length is far smaller than the ion
gyroradius, i.e., $\lambda_D\ll\rho_i$. For the present work, we
focus on the one-dimensional (1D) physics along the field line and
thus ${\bf b} \cdot {\bf \nabla} =
\partial_{l}$.

For initial value approach, we can define
\begin{equation}\label{eq:gk_g}
g_s\equiv h_s-\frac{q_{s}}{T_{s}} F^{s}_{0}J_{0} \phi, ~{\rm
i.e.},~f_s=-\frac{q_{s}}{T_{s}} F^{s}_{0} (1-J_0^2)\phi + J_{0}
g_{s},
\end{equation}
and Eq.(\ref{eq:gk_dh}) changes to (with also $\omega=i\partial_t$)
\begin{eqnarray}\nonumber
(\partial_t  +  v_{\parallel} \partial_{l} ) g_{s}
&=&-i\omega_{Ds}g_{s} -i(\omega_{Ds} - \omega_{*s}^{T}
  )\frac{q_{s}}{T_{s}} F^{s}_{0}J_{0} \phi \\\label{eq:gk_dg}
  &&-v_{\parallel} \frac{q_{s}}{T_{s}} F^{s}_{0}[J_{0} \partial_{l} \phi-J_{1} \partial_{l}(k_\perp\rho_s)
  \phi],
\end{eqnarray}
where we have used $J'_0=-J_1$. For only one species of ion, i.e.,
$s=i,e$ and $q_i=-q_e=e$, Eq. (\ref{eq:gk_n}) can be rewritten as
\begin{eqnarray}\label{eq:gk_n2}\nonumber
&&-\frac{q_{i}n_i}{T_{i}} (1-\Gamma_{0i})\phi +  \int J_{0i} g_{i}
d^{3} v = \\
&&-\frac{q_{e}n_e}{T_{e}} (1-\Gamma_{0e})\phi + \int J_{0e} g_{e}
d^{3} v,
\end{eqnarray}
where $\Gamma_{0s}\equiv I_0(b_s)e^{-b_s}$, with $I_0$ the modified
Bessel function, $b_s=\Big(\frac{k_\perp v_{ts}}{\Omega_s}\Big)^2$
and $v_{ts}=\sqrt{T_s/m_s}$. We have solved the above gyrokinetic
system Eqs.(\ref{eq:gk_dg}) and (\ref{eq:gk_n2}) in Z-pinch with
only passing particles using MGK code in Ref.\cite{Xie2017a}, where
we can assume $v_\parallel$ and $v_\perp$ to be constant along field
line. However, to study the physics in dipole configuration, we must
treat the particles trapping, i.e., the variation of $v_\parallel$
and $v_\perp$ along the field line.

\section{Dipole Equilibrium Operators}\label{sec:dip_op}

For idea point dipole or current loop ring dipole, the equilibrium
magnetic field ${\bf B}$ is symmetric in toroidal direction, thus we
can write
\begin{eqnarray*}
  {\bf B} &=& \nabla \zeta \times \nabla \psi=\nabla\chi, \\
  {\bf k}_{\perp} &=& k_{\psi} \nabla \psi  + k_{\zeta } \nabla \zeta ,
\end{eqnarray*}
where $\zeta$ is azimuthal (toroidal) angle, and $\psi$ is flux
surface function (radial). The above equilibrium can be a good
equilibrium model for plasma at low $\beta$, where $\beta$ is the
ratio of plasma pressure and magnetic pressure.

In this section, we summarize the derivations of the operators in
our simulation model for ring dipole configuration and leave the
detailed derivations using point dipole as example in the Appendix
\ref{sec:apd_point_op}. We firstly define torus coordinates
$(r,\theta,\phi)$ with $R=a+r\cos\theta$, and flux coordinates
$(\psi,\chi,\zeta)$ with $\psi=\psi(r,\theta)$,
$\chi=\chi(r,\theta)$ and $\zeta=\phi$. And we define
$\xi=\theta\in[-\pi,\pi]$, along the field line
\begin{equation}\label{eq:dldtheta}
 dl=r_0\kappa(\xi) d\xi,
\end{equation} where we take $r_0=\frac{R(\theta=0)-R(\theta=\pi)}{2}$, and we have
\begin{equation}\label{eq:dl}
\frac{\partial}{\partial l}=\frac{\partial}{\partial \xi}
\frac{\partial \xi}{\partial
l}=\frac{1}{r_0\kappa}\frac{\partial}{\partial
 \xi}.
\end{equation}
For point dipole, $\kappa=\cos\xi(1+3\sin^2\xi)^{1/2}$,
$\xi=\pi/2-\theta$ and $\theta$ is spherical coordinate. For
Z-pinch, $\kappa=1$. Considering $\epsilon$ and $\mu$ conserved,
pitch angle $\lambda\equiv\mu B_0/\epsilon$,
$v_\parallel=\sqrt{2(\epsilon-\mu B)}=v\sqrt{1-\lambda B/B_0}$,
$v=\sqrt{2\epsilon}$, $v^2=v_\parallel^2+v_\perp^2$, $B=B_0f(\xi)$
and $v_\parallel=dl/dt$, we have
\begin{equation}\label{eq:dxidt}
 \frac{d\xi}{dt}=\frac{v_\parallel}{r_0\kappa}=\frac{\pm
 v\sqrt{y}}{r_0\kappa},
\end{equation}
with $y=1-f(\xi)\lambda$. For point dipole,
$f(\xi)=\frac{\sqrt{1+3\sin^2\xi}}{\cos^6\xi}$. For Z-pinch, $f=1$.

\begin{figure}
 \centering
  \includegraphics[width=8.5cm]{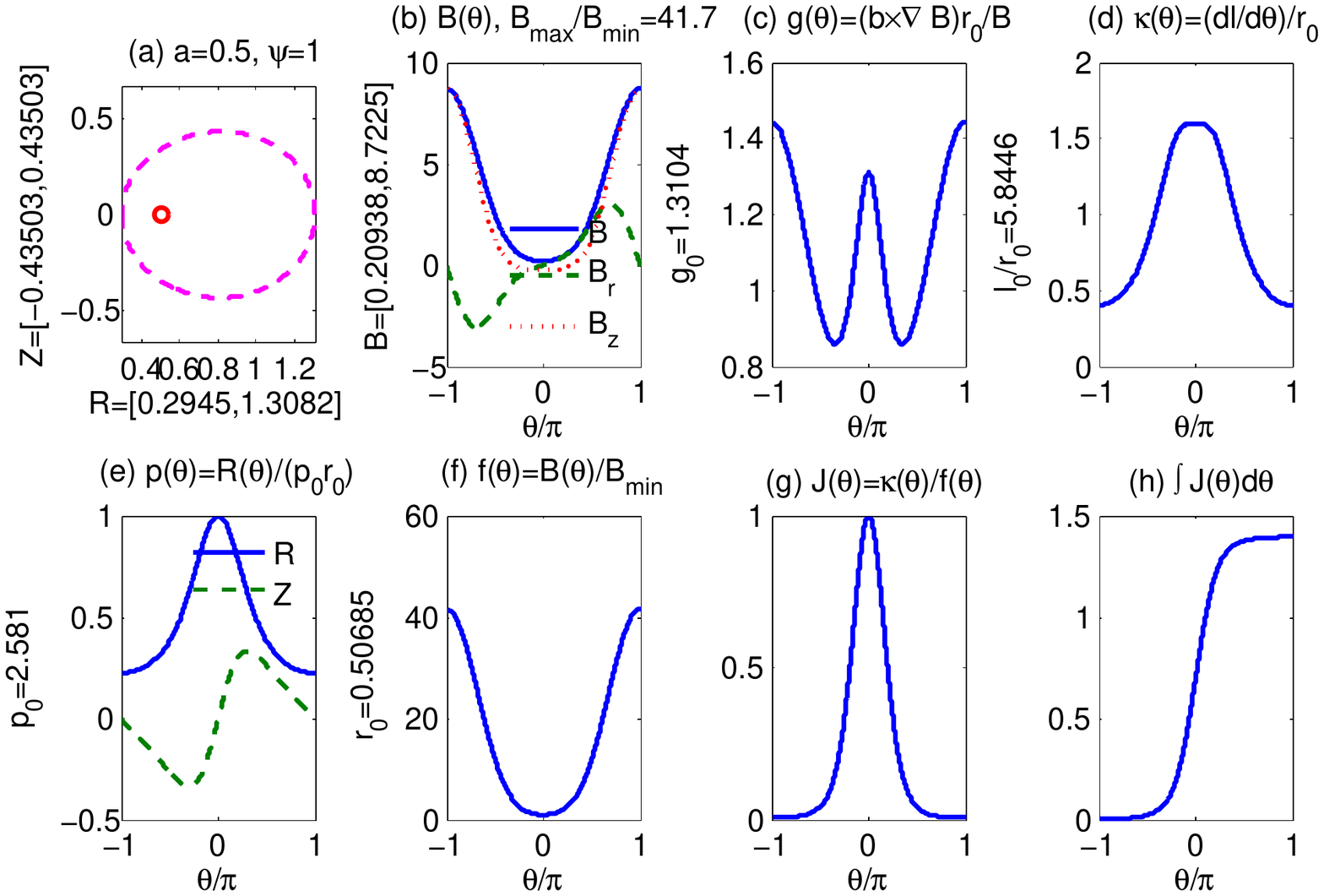}\\
  \caption{Configuration functions along a field line in ring dipole case.
  }\label{fig:ftheta}
\end{figure}

Taking ${\bf k}_{\perp} = k_{\psi} \nabla \psi  + k_{\zeta } \nabla
\zeta $, $k_\perp=|{\bf k}_{\perp}|$, $k_\psi=0$ for ring dipole,
and using ${\bf B}= \nabla \zeta \times \nabla \psi=\nabla\chi$ and
$\nabla \zeta= \frac{1}{R}{\bm \hat e}_\phi$, we obtain along a
field line
\begin{equation*}
    k_\perp^2=k_{\zeta }^2 \nabla \zeta\cdot \nabla
    \zeta=k_{\zeta
    }^2\frac{1}{r_0^2p_0^2p^2}=k_t^2\frac{1}{r_0^2p^2},
\end{equation*}
where we have taken $R=r_0p_0p(\xi)$, {\color{red}i.e.,
$p(\xi)\equiv R/(r_0p_0)$,} and $k_t\equiv k_\zeta/p_0$.
We introduce a constant $p_0$ to make the configuration
function $p(\xi=0)=1$, and hence the notations can be convenient for
different configurations. An example of $p(\xi)$ for ring dipole is
shown in Fig.\ref{fig:ftheta}(e), where $R$ and $Z$ are the
cylindrical coordinate $(R,\zeta,Z)$ normalized by $R(\theta=0)$.
For Z-pinch, we have $p_0=a/r_0$ and $p=1$. We can also readily
obtain $\rho_{s} =\frac{v_{\perp}}{\Omega_{s}}= \frac{\sqrt{2\mu
B}}{q_s
  B/m_sc}=\frac{\sqrt{2\mu
  B_0}}{\Omega_{s0}}\sqrt{\frac{B_0}{B}}=\frac{v\sqrt{\lambda}}{\Omega_{s0}}\frac{1}{\sqrt{f(\xi)}}$.
Thus $k_\perp\rho_{s} =
k_t\frac{v\sqrt{\lambda}}{r_0\Omega_{s0}}\frac{1}{z}$, with
$z(\xi)=p(\xi)\sqrt{f(\xi)}$, and $\partial_\xi( k_\perp\rho_s) =
k_t\frac{v\sqrt{\lambda}}{r_0\Omega_{s0}}\partial_\xi(z^{-1})=-k_t\frac{v\sqrt{\lambda}}{r_0\Omega_{s0}}z^{-2}\partial_\xi
z$, and $b_s=\Big(\frac{k_\perp v_{ts}}{\Omega_s}\Big)^2=
    \frac{k_t^2v_{ts}^2}{r_0^2\Omega_{s0}^2}\frac{1}{p^2f^2}$.

Since $n_0=n_0(\psi)$ and $T_0=T_0(\psi)$, we can have
\begin{eqnarray*}
    \nabla F_0&=&\nabla\Big[n_0(\frac{m}{2\pi T_0})^{3/2}e^{-m\epsilon/T_0}\Big]\\
    &=&\Big\{\frac{\nabla n_0}{n_0}+\frac{\nabla T_0}{T_0}\Big[\frac{m\epsilon}{T_0}-\frac{3}{2}\Big]\Big\}F_0\\
    &=&-\nabla\psi \Big\{L_n^{-1}+L_T^{-1}\Big[\frac{m\epsilon}{T_0}-\frac{3}{2}\Big]\Big\}F_0\\
    &=&\frac{\nabla n_0}{n_0}\Big\{1+\eta_s\Big[\frac{m\epsilon}{T_0}-\frac{3}{2}\Big]\Big\}F_0,
\end{eqnarray*}
where $T_s=m_sv_{ts}^2$, $L_n^{-1}\equiv-\frac{\partial \ln
n_0}{\partial \psi}$, $L_T^{-1}\equiv-\frac{\partial \ln
T_0}{\partial \psi}$ and $\eta_s=\frac{L_n}{L_{T_s}}$. And by
further using
\begin{eqnarray*}
    ({\bf k_\perp}\times {\bf b})\cdot\nabla n_0&=&\frac{1}{B}[(k_{\psi} \nabla \psi  +
k_{\zeta } \nabla \zeta)\times \nabla \chi]\cdot \nabla \psi \frac{\partial n_0}{\partial \psi}\\
    &=&\frac{1}{B}[
k_{\zeta } \nabla \zeta\times \nabla \chi]\cdot \nabla \psi
\frac{\partial n_0}{\partial \psi}\\
    &=&k_{\zeta }B
\frac{\partial n_0}{\partial \psi},
\end{eqnarray*}
we obtain
\begin{eqnarray*}
  \omega_{*s}^T &=& \frac{{\bf k}_{\perp} \times {\bf b}
\cdot {\bf \nabla} F^{s}_{0} }{
-\Omega_{s}F^{s}_{0\epsilon}}=\frac{{\bf k}_{\perp} \times {\bf b}
\cdot {\bf \nabla} F^{s}_{0} }{ m_s\Omega_{s}F^{s}_{0}/T_s}\\
&=& \frac{k_{\zeta }B \frac{\partial n_0}{\partial
\psi}\frac{1}{n_0}\Big\{1+\eta_s\Big[\frac{m\epsilon}{T_0}-\frac{3}{2}\Big]\Big\}}{m_s\Omega_{s}/T_s}\\
&=&
-\omega_{*s}\Big\{1+\eta_s\Big[\frac{m\epsilon}{T_0}-\frac{3}{2}\Big]\Big\},
\end{eqnarray*}
where $\omega_{*s}=\frac{k_\zeta c T_s}{q_s L_{ns}}$.

The gradient drift ${\bm v}_g$, curvature drift ${\bm v}_c$, and
total drift ${\bm v}_d={\bm v}_g+{\bm v}_c$ are
\begin{equation}
  {\bm v}_g=\frac{1}{m\Omega_s}\mu {\bm b}\times\nabla B=-\frac{v_\perp^2}{2\Omega_sB}({\bm b}\times\nabla B){{\hat
e}_\phi},
\end{equation}
\begin{equation}
  {\bm v}_c=\frac{1}{\Omega_s}v_\parallel^2\nabla\times {\bm b}=-\frac{v_\parallel^2}{\Omega_sB}({\bm b}\times\nabla B){{\hat
e}_\phi},
\end{equation}
\begin{equation}
  {\bm v}_d=-\frac{{\bm b}\times\nabla B}{B\Omega_s}(v_\parallel^2+\frac{1}{2}v_\perp^2){{\hat
  e}_\phi}
  =-\frac{g}{r_0\Omega_{s0}f}(v_\parallel^2+\frac{1}{2}v_\perp^2){{\hat
  e}_\phi},
\end{equation}
where we have defined $({\bm b}\times\nabla B)/B=g(\theta)/r_0$,
which would be calculated numerically. Note that $\nabla\times {\bm
b}=\nabla\times \frac{\bm B}{B}=\frac{1}{B}\nabla\times {\bm
B}+\Big(\nabla\frac{1}{B}\Big)\times {\bm B}=\frac{{\bm b}}{B}\times
\nabla B$, where we have used $\nabla\times {\bm B}=0$ for vacuum
field. And thus we obtain
\begin{equation}
  \omega_{Ds}={\bf k_\perp}\cdot{\bm v}_d
  =\frac{-k_tg}{r_0^2\Omega_{s0}fp}(v_\parallel^2+\frac{1}{2}v_\perp^2)
  =\frac{-\omega_{d0}g}{fp}\frac{(1+y)}{2}v^2,
\end{equation}
where $\omega_{d0}=\frac{k_t}{r_0^2\Omega_{s0}}$. For point dipole,
$g(\theta)=\frac{3(\cos^2\theta+1)}{\sin\theta(1+3\cos^2\theta)^{3/2}}$,
$r=r_0\sin^2\theta$. For Z-pinch, $g=1$.

For a typical ring dipole parameter, the corresponding $B$, $f$,
$\kappa$, $g$ and $p$ on $\theta$ are shown in Fig.\ref{fig:ftheta},
where the $\bm B$ field is calculated using the elliptic functions
based on Appendix \ref{sec:apd_ring_eq}. This typical ring dipole
configuration will be used in later simulations in
Sec.\ref{sec:sim_result}.

Treating electron and ion using the same kinetic equation, with
$q_i=-q_e=e$, $n_{i0}=n_{e0}$, $\tau_e=T_e/T_i$, and defining the
$\delta f$ weight $w=g/F_0$, the kinetic equation
Eq.(\ref{eq:gk_dg}) can be rewritten as
\begin{equation}\label{eq:gk_dxis}
\frac{d\xi_s}{dt} = \frac{v_\parallel}{r_0\kappa},
\end{equation}
\begin{eqnarray}\label{eq:gk_dws}
 \frac{dw_s}{dt}&=& -i\omega_{Ds}w_s -i(\omega_{Ds} -
\omega_{*s}^{T}
  )\frac{q_s}{T_{s}} J_{0} \phi \\\nonumber
  &&-v_{\parallel} \frac{q_s}{T_{s}} \frac{1}{r_0\kappa}[J_{0} \partial_{\xi} \phi-J_{1}
  \partial_{\xi}(k_\perp\rho_s)
  \phi],
\end{eqnarray}
and the quasi-neutrality Eq. (\ref{eq:gk_n2}) does not change.

Considering $\tau=\tau_e=T_e/T_i$, $v_{te}=v_{ti}\sqrt{\tau
m_i/m_e}$, $\Omega_e=\frac{q_em_i}{q_im_e}\Omega_i$,
$\rho_{te}=\rho_{ti}\frac{q_i}{q_e}\sqrt{\tau\frac{m_e}{m_i}}$, and
defining $k_s\equiv k_\perp\rho_{ts}$, we normalize the equations by
$v_0$ and $R_0$. We assume hot ion, and take $v_0=v_{ti}$,
$R_0=r_0$. Thus, we have: length $L\to L/r_0$, velocity $v\to
v/v_{0}$, time $t_0=r_0/v_{0}$, $t\to t/t_0$, frequency
$\omega_0=v_{0}/r_0$, $\rho_{ti}=v_{0}/\Omega_{0i}$,
$\epsilon_n\equiv-\frac{1}{B_0p_0r_0^2}\Big(\frac{\partial \ln
n_0}{\partial \psi}\Big)^{-1}=L_n/(B_0p_0r_0^2)$, $\phi\to
e\phi/T_i$. The normalized variables are $\rho_{ti}\to
\rho_{ti}/r_0=\omega_0/\Omega_{0i}$,
$\rho_{te}=\rho_{ti}\frac{q_i}{q_e}\sqrt{\tau\frac{m_e}{m_i}}$,
$F_{0s}=(\frac{1}{2\pi v_{ts}^2})^{3/2}e^{-v^2/2v_{ts}^2}$,
$v_{ti}\to1$, $v_{te}\to\sqrt{\tau m_i/m_e}$, and wavevectors
$k_\perp\to k_\perp r_0$, $k_\zeta\to k_\zeta$, $k_\psi\to k_\psi
B_0r_0^2$. And hence
\begin{equation*}
    k_\perp^2=k_t^2/p^2,~~~
  k_\perp\rho_{s} = \frac{v}{v_{ts}}\sqrt{\lambda}k_t\rho_{ts}z^{-1},
\end{equation*}
\begin{equation*}
  \partial_\xi( k_\perp\rho_s) = -\frac{v}{v_{ts}}\sqrt{\lambda}k_t\rho_{ts}z^{-2}\partial_\xi
  z,~~~
    b_s=\rho_{ts}^2k_t^2\frac{1}{p^2f^2}.
\end{equation*}
Normalized diamagnetic drift frequency is
\begin{equation*}
  \omega_{*s}^T =
-\omega_{*s}\Big\{1+\eta_s\Big[\frac{v^2}{2v_{ts}^2}-\frac{3}{2}\Big]\Big\}=
-\omega_{s0}\Big\{\kappa_n+\kappa_T\Big[\frac{v^2}{2v_{ts}^2}-\frac{3}{2}\Big]\Big\},
\end{equation*}
where $\omega_{*i}=k_t \omega_0\rho_{ti}/\epsilon_{n}\to k_t
\rho_{ti}/\epsilon_{n}$,
$\omega_{*e}=\frac{q_i}{q_e}\tau\omega_{*i}$,
$\kappa_n=\epsilon_n^{-1}$ and $\kappa_T=\eta_s\epsilon_n^{-1}$.
Normalized curvature drift frequency is
\begin{equation*}
  \omega_{Ds}
  =-\omega_{ds0}\frac{g}{fp}\frac{(1+y)}{2}\frac{v^2}{v_{ts}^2},
\end{equation*}
where $\omega_{di0}=k_t\omega_0\rho_{ti}\to k_t\rho_{ti}$ and
$\omega_{de0}=\tau\frac{q_i}{q_e}\omega_{di0}$.

The final gyrokinetic system changes to
\begin{equation}\label{eq:gk_dxi2}
 \frac{d\xi_s}{dt} = \frac{v_\parallel}{\kappa},
\end{equation}
\begin{eqnarray}\label{eq:gk_dw2}
  \frac{dw_s}{dt}&=& -i\omega_{Ds}w_s -i(\omega_{Ds} -
\omega_{*s}^{T}
  ) \frac{q_s}{T_s}J_{0} \phi \\\nonumber
  &&-v_{\parallel} \frac{1}{\kappa}\frac{q_s}{T_s}[J_{0} \partial_{\xi} \phi-J_{1}
  \partial_{\xi}(k_\perp\rho_s)
  \phi],
\end{eqnarray}
\begin{equation}\label{eq:gk_ni2}
\Big(1+\frac{1}{\tau_e}-\Gamma_{0i}-\frac{1}{\tau_e}\Gamma_{0e}\Big)\phi=
\int J_{0i} g_{i} d^{3} v-\int J_{0e} g_{e} d^{3} v.
\end{equation}
To avoid the sign change of $v_\parallel$ at the turning point, we
add an extra equation to calculate $v_\parallel$
\begin{equation}\label{eq:gk_dvpar2}
\frac{dv_\parallel}{dt} =
 -\frac{v^2\lambda}{2r_0}\frac{1}{\kappa}\frac{df}{d\xi}\to-\frac{v^2\lambda}{2}\frac{1}{\kappa}\frac{df}{d\xi},
\end{equation}
which can be derived from Eq.(\ref{eq:dxidt}) via $\mu$ and
$\epsilon$ conservation. For point dipole, $\frac{dv_\parallel}{dt}
=-\frac{v^2\lambda}{2}\frac{3\sin\xi(3+5\sin^2\xi)}{\cos^8\xi(1+3\sin^2\xi)}$,
which can also be derived via mirror force $F=-\mu\nabla B$, i.e.,
$F_\parallel=-\mu (\partial_\chi B) |{\bm e}^{\chi}|$.
Eqs.(\ref{eq:gk_dxi2})-(\ref{eq:gk_dvpar2}) are our final equations
to solve for local electrostatic drift mode in dipole. Standard
$\delta f$ particle-in-cell (PIC) approach
\cite{Parker1993,Xie2017a} is used in this work to do our
simulation. If we use adiabatic electron model, i.e., $h_e=0$ (not
$g_e=0$), the only change is the quasi-neutrality
Eq.(\ref{eq:gk_ni2}), which should be changed to
\begin{equation}\label{eq:gk_ni_adab}
\Big(1+\frac{1}{\tau_e}-\Gamma_{0i}\Big)\phi= \int J_{0i} g_{i}
d^{3} v.
\end{equation}
In the latter part, we will check whether the adiabatic electron
model is valid for dipole simulations.

One should also note the slight difference of the definitions of our
notations between ring dipole and point dipole. These differences
mainly come from the coordinate system: for ring dipole we calculate
our variables from torus coordinate; whereas for point dipole we
calculate them from spherical coordinate. Thus, for examples, if we
reduce $a\to0$ in ring dipole case, we obtain the curvature radius
at $\xi=0$ to be $R_c^{\rm ring-dipole}=2R/3$, which is not
$R_c^{\rm point-dipole}=R/3$ as in the point dipole case. And for
Z-pinch $R_c^{\rm Z-pinch}=R$. That is, if we want to quantitatively
compare the results in ring dipole and point dipole, we should keep
in mind the normalization $R$ is $R^{\rm point}\sim2R^{\rm ring}$.
This difference will affect $\kappa_n$, $\kappa_T$, $\omega_D$ and
so on. In our normalization in this work, $\omega_{*i}<0$ and
$\omega_{*e}>0$, and thus positive real frequency means the mode
propagates in electron diamagnetic direction and negative real
frequency means the mode propagates in the ion diamagnetic
direction.

\section{Zero-dimensional Dispersion relation}\label{sec:dr}

In ideal dipole field, all particles are trapped particles and the
simplest dispersion relation is replacing $\omega_D$ by bounce
average $\langle\omega_D\rangle_b$, and setting $\langle
J_0(k_\perp\rho)\phi\rangle_b=\langle J_0(k_\perp\rho)\rangle_b\phi$
by assuming $k_\parallel=0$. We have {\small
\begin{equation}\label{eq:entropy_DR0}
    \sum_{s=e,i}\int dv^3[1-\omega_*^T(v)/\omega]\partial_\epsilon F(v)\Big\{\phi-
    \frac{J_0\oint J_0\phi dl/v_\parallel}{\oint[1-\omega_D(v)/\omega]dl/v_\parallel}\Big\}=0,
\end{equation}}
where $F(v)$ is Maxwellian isotropic equilibrium distribution
function. This yields the normalized final dispersion relation
{\small
\begin{equation}\label{eq:entropy_DR}
    D(\omega,k)=\sum_{s=e,i}\frac{1}{T_s}\Big\{1-\int dv^3\frac{[\omega-
    \omega_*^T(v)]J_0\langle J_0\rangle_b}{[\omega-k_{zs} v_\parallel
    -\langle\omega_D(v)\rangle_b]}e^{-\frac{v^2}{2}}\Big\}=0.
\end{equation}}
where $\int
dv^3=\frac{1}{\sqrt{2\pi}}\int_{-\infty}^{\infty}dv_\parallel\int_0^{\infty}v_\perp
dv_\perp $, and we have artificially added back the $k_\parallel$
term to make it more general, with $k_{zi}=k_\parallel R$ and
$k_{ze}=k_\parallel R\sqrt{\tau m_i/m_e}$. Eq.(\ref{eq:entropy_DR})
can be seen as an extension of the one in
Refs.\cite{Rutherford1968,Kesner1998}. The above dispersion relation
Eq.(\ref{eq:entropy_DR}) is similar to the one in
Z-pinch\cite{Ricci2006,Xie2017a}, except several bounce average
terms. The bounce average $\langle \omega_D\rangle_b$ and $\langle
J_0\rangle_b$ can be found at Appendix \ref{sec:apd_bounce_avg}.
Interpolation or fitting can be used to speed up the numerical
calculation. The details of the root finding method can be found at
Ref.\cite{Xie2017a}. One should also note that it is not easy to do
the bounce average accurately especially when considering
$k_\perp\rho_i$ not small in $J_0$.
Refs.\cite{Kesner2002,Helander2016} only discussed the $J_0\to1$
limit. As will be shown in Fig.\ref{fig:plt_w_k_dip}, if we modify
the term $J_0\langle J_0\rangle_b $ to $J_0^2$, $\langle
J_0\rangle_b^2$ or $\langle J_{0}^2\rangle _b$, the solutions will
change quantitatively but not qualitatively, i.e., the essential
physics is the same.

\section{Particle-in-cell approach}\label{sec:pic}

\begin{figure}
 \centering
  \includegraphics[width=8.0cm]{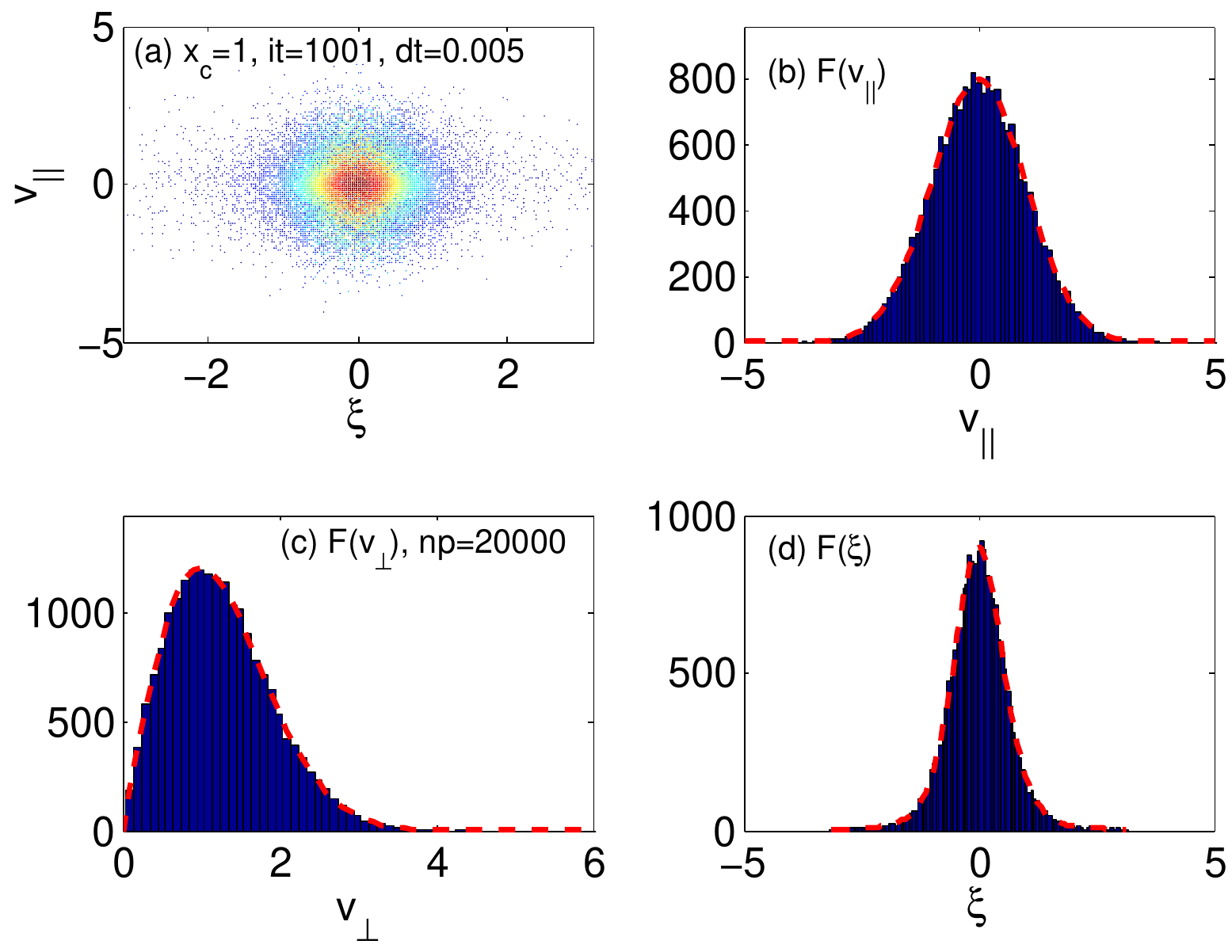}\\
  \caption{After $nt=1000$ time steps, the distribution function $F(\theta,v_\parallel,v_\perp,t)$
   (the bar charts) still agrees with the initial loading
  $F_0=F(\theta,v_\parallel,v_\perp,t=0)$ (the red dash line) in ring dipole case.
  }\label{fig:gkd1d_f0}
\end{figure}

The main steps of $\delta f$ particle-in-cell model
\cite{Parker1993} used here are summarized in Ref.\cite{Xie2017a}.
However, in this work we should carefully treat the non-uniform
magnetic field, which strongly affects the particle loading to keep
the equilibrium distribution function $F_0$ constant with time,
i.e., $\partial F_0/\partial t=0$. In our simulation model, we load
the velocity using Gaussian random number as in Ref.\cite{Xie2017a},
i.e., Maxwellian $v_\parallel=randn(np)\times v_{t}$ and
$v_\perp=\sqrt{randn(np)^2+randn(np)^2}\times v_{t}$. Here, $n_p$ is
particle number for one species, $j=1,2,\cdots,n_p$ is particle
index, and $randn()$ generates normal distribution
$F_0=\frac{1}{\sqrt{2\pi}}e^{-x^2/2}$. The initial spatial position
$\xi_j$ should be loaded according to the flux-tube volume, i.e.,
$F_0(\xi)\propto J(\theta)=\kappa(\theta)/f(\theta)$, where
$J(\theta)$ can be seen as Jacobian metric. For point dipole,
$J(\xi)=\cos^7\xi$. We use acceptance-rejection method to generate
this nonuniform loading of $\xi_j$. We note that the particle
loading approaches in Refs.\cite{Dettrick2003,Bierwage2008} are
different, which are more complicated. We have verified our approach
that the equilibrium distribution function indeed remains unchanged
with time, as shown in Fig.\ref{fig:gkd1d_f0}.

We use periodic boundary condition for field, i.e.,
$\phi(n_g+1)=\phi(1)$, where $n_g$ is the field grid number for
spatial coordinate $\xi$. The particles are also treated
periodically, i.e., if a particle passes one boundary, we let it
enter the simulation domain at another boundary. The simulation box
is $\xi/\pi\to[-x_c,x_c]$. That is, we do not need to simulate the
whole field line. If we set $x_c\to0$, the simulation results should
reduce to the slab case and should agree with the dispersion
relation accurately. By varying $x_c$ we examine how the results
change from Z-pinch configuration to dipole configuration, i.e.,
$x_c\to0$ for Z-pinch case and $x_c=1$ (or $x_c\to0.5$) for ring (or
point) dipole case.

One difficulty of the present simulation model is to study the
$k_\perp\rho_i\ll1$ modes at large $B$ field region, due to the term
$G_{\rm
coef}=\Big(1+\frac{1}{\tau_e}-\Gamma_{0i}-\frac{1}{\tau_e}\Gamma_{0e}\Big)\simeq
b_i^2\to0$ in the field equation, especially for $\xi$ around the
simulation edge ($|\xi|\to\pi/2$). We take point dipole for example:
The magnetic field at $\xi_c$ is $f(\xi_c)$; The ratio of particles
at $|\xi|>\xi_c$ is $N_{\rm
ratio}=1-\int_{-\xi_c}^{\xi_c}J(\xi)d\xi/\int_{-\pi/2}^{\pi/2}J(\xi)d\xi$.
Some typical values are listed in Table.\ref{tab:fxi}. We can see
that only $N_{\rm ratio}\sim10^{-5}$ particles exist at
$|\xi|>0.4\pi$ which is difficult to represent the density integral
$\int J_{0i} g_{i} d^{3} v$ accurately and the coefficient $G_{\rm
coef}\sim10^{-4}$ is very small if we take $k_\perp\rho_i\sim1$.
Thus, the electrostatic potential $\phi$ calculated from the
quasi-neutrality Eq.(\ref{eq:gk_ni2}) will have large numerical
error at larger $|\xi|$. At this stage, we have not fully resolved
this difficulty but use large particle number $n_p$ and adjust $x_c$
to overcome it.

\begin{table}[!h]
\tabcolsep 5mm \caption{Numerical difficulty for small
$k_\perp\rho_i$ due to the strong magnetic at edge ($|\xi|\to\pi/2$)
in point dipole.}\label{tab:fxi}
\begin{center}
\begin{tabular}{c|ccc}
\hline\hline $\xi_c/\pi$ & $B/B_0$ & $b_i^2$ & $N_{\rm ratio}$
\\\hline
0 & 1 & 1 & 1 \\
0.05 & 1.116 & 0.8649 & 0.6661 \\
0.1 & 1.533 & 0.5752 & 0.3850 \\
0.15 & 2.542 & 0.3092 & 0.1875 \\
0.2 & 5.090 & 0.1377 & 0.07394 \\
0.25 & 12.65 & 0.05 & 0.0222 \\
0.3 & 41.74 & 0.0139 & 4.59e-3 \\
0.35 & 210.0 & 2.59e-3 & 5.40e-4 \\
0.4 & 2.21e4 & 2.35e-4 & 2.37e-5 \\
0.45 & 1.35e5 & 3.73e-6 & 9.90e-8 \\
0.5 & $\infty$ & 0 & 0 \\\hline\hline
\end{tabular}
\end{center}
\end{table}

\section{Benchmark and basic features of the modes}\label{sec:bech}
\begin{figure}
 \centering
  \includegraphics[width=8.0cm]{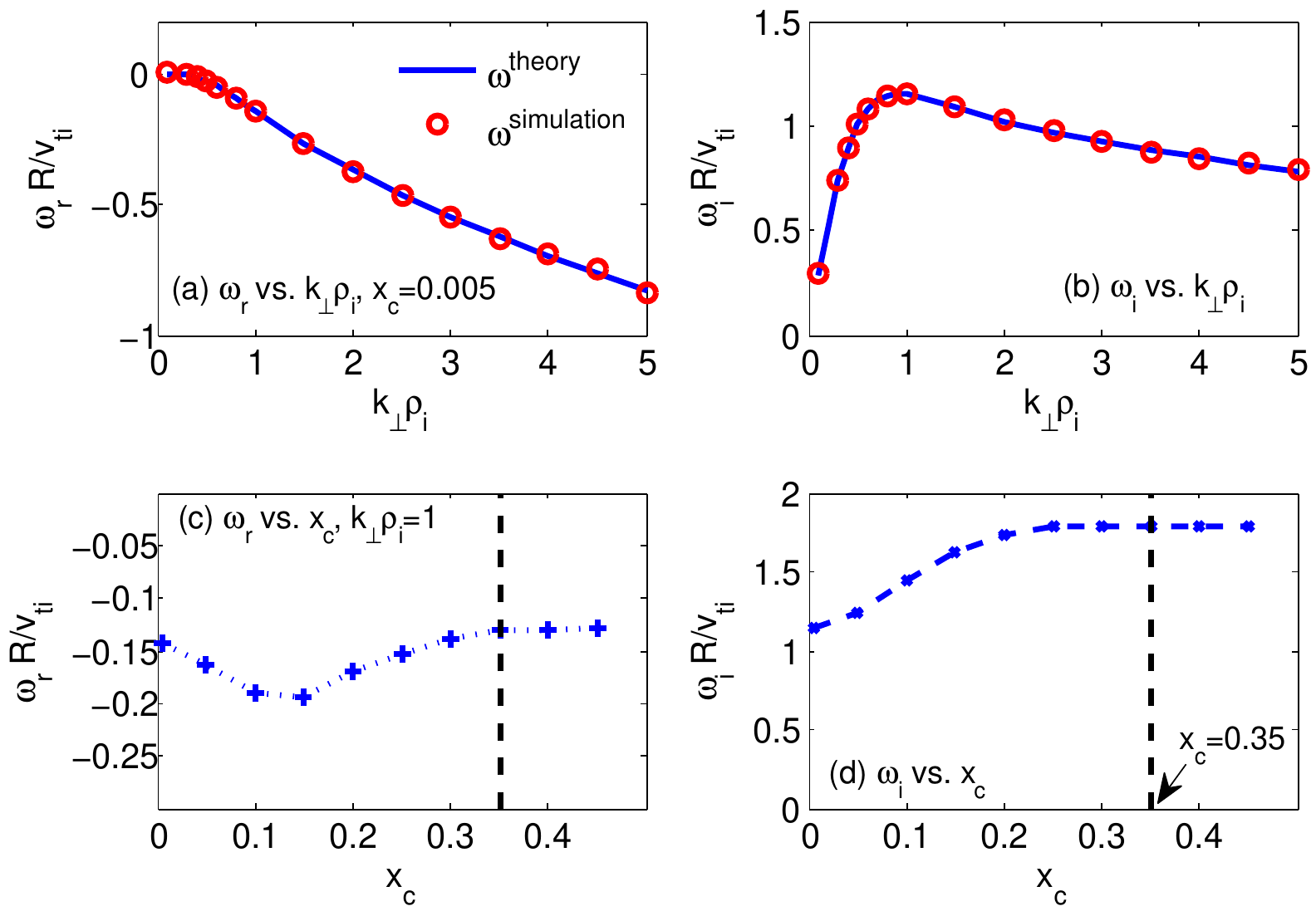}\\
  \caption{Benchmark of the simulation model in point dipole configuration: (a\&b) Scanning $\omega$ vs. $k_\perp\rho_i$ with $x_c=0.005$,
  and comparing with the slab dispersion relation solution; (c\&d) Scanning $\omega$ vs. $x_c$.}\label{fig:bench_01}
\end{figure}

We firstly benchmark our simulation model with slab (Z-pinch) case
by setting $x_c=0.005$, with $\kappa_n=5$, $\kappa_T=0$,
$n_p=3\times10^5$ and $n_g=4$. The other default parameters are
$\tau=1$, $k_\psi=0$ and $m_i/m_e=1836$. If not specialized,
hereafter in the figures $k_\perp\rho_i$ represents $k_t\rho_{ti}$,
i.e., the normalized perpendicular wavevector with $k_\psi=0$ at
$\xi=0$. The simulation result in point dipole configuration is
shown in Fig.\ref{fig:bench_01}(a\&b), where the theoretical result
is calculated via Eq.(\ref{eq:entropy_DR}) by setting $\omega_d^{\rm
point-dipole}=3\omega_d^{\rm Z-pinch}$ and without bounce average.
We can find that the simulation result agrees with dispersion
relation solution with error less than $1\%$.

Figure \ref{fig:bench_01}(c\&d) shows a scanning of $\omega$ vs.
$x_c$ in point dipole configuration. We can find that the $\omega$
changes little for $x_c\geq0.3$, where $f(0.3\pi)\sim41.7\gg1$,
since few particles exist at $\xi\to\pm\pi/2$ due to strong magnetic
field $B(\xi\to\pm\pi/2)/B_0\gg1$ in point dipole. Due to this
convergence of $x_c$, we will set $x_c=0.35$ as default in our
simulation in point dipole. And this can also avoid the difficulty
on suppressing noise at the boundary when $x_c\to0.5$, e.g., there
exists less than $10$ particles at $|\xi|>0.45\pi$ even for
$n_p=10^{8}$.



\begin{figure}
 \centering
  \includegraphics[width=8.0cm]{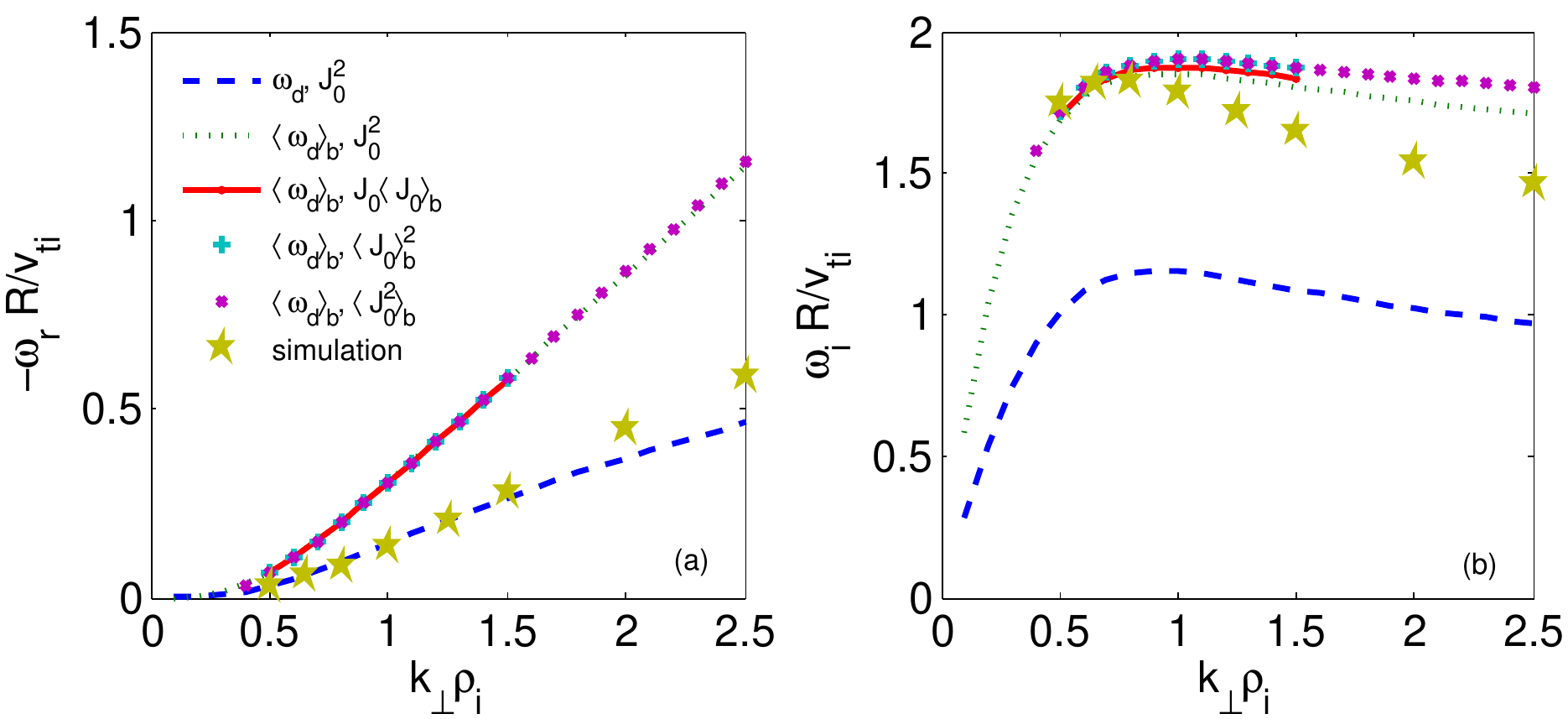}\\
\caption{Comparisons of point dipole simulation result with
dispersion relation solutions with different types of bounce
averages of $\omega_D$ and $J_0$. } \label{fig:plt_w_k_dip}
\end{figure}

\begin{figure}[htbp]
\centering
\includegraphics[width=8cm]{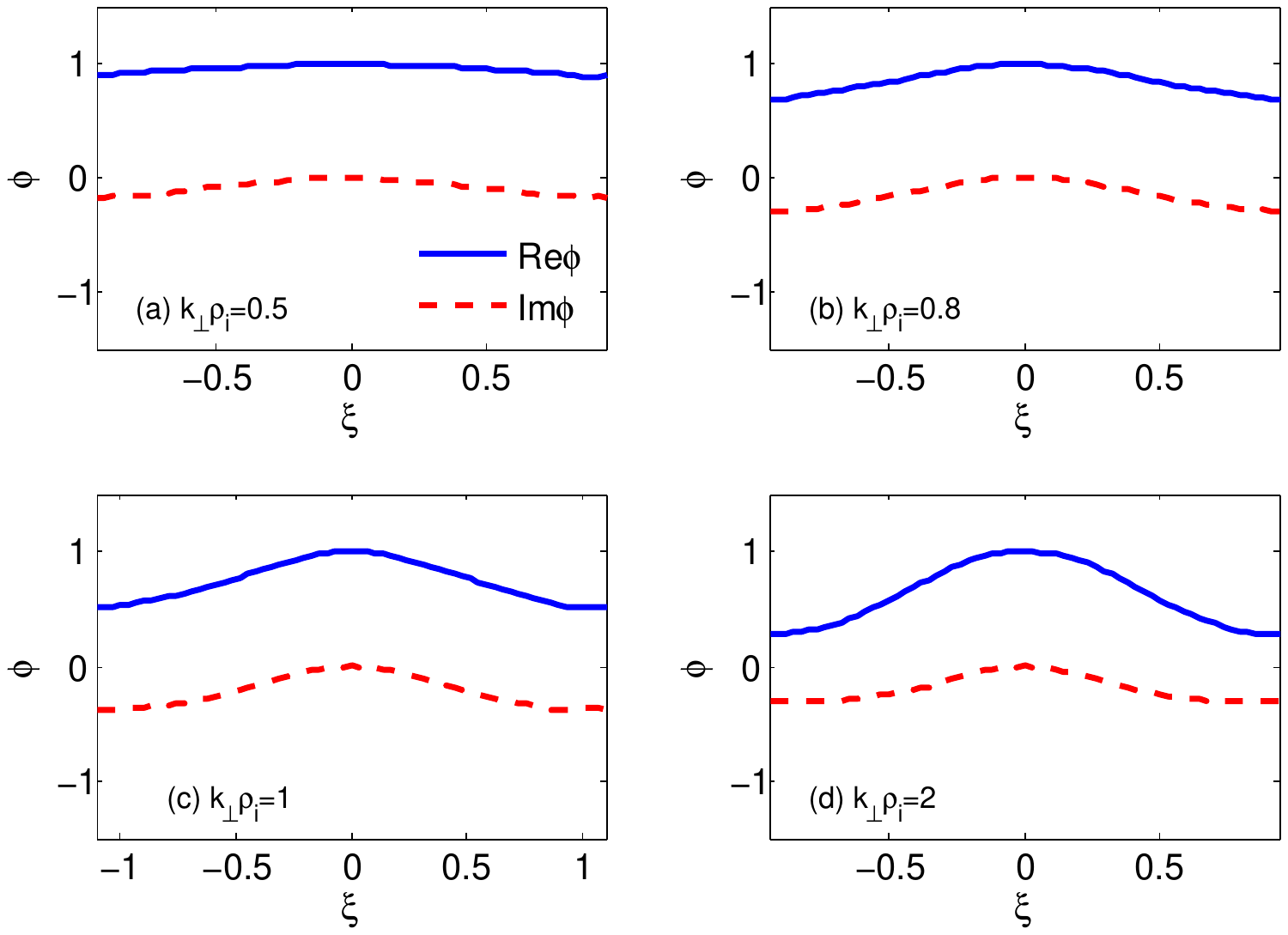}
\caption{Mode structures for $\kappa_n=5.0$, $\kappa_T=0.0$ at point
dipole with $k_\perp\rho_i=0.5$, $0.8$, $1.0$ and $2.0$
respectively.} \label{fig:phi_xi_point_k}
\end{figure}

\begin{figure}[htbp]
\centering
\includegraphics[width=8cm]{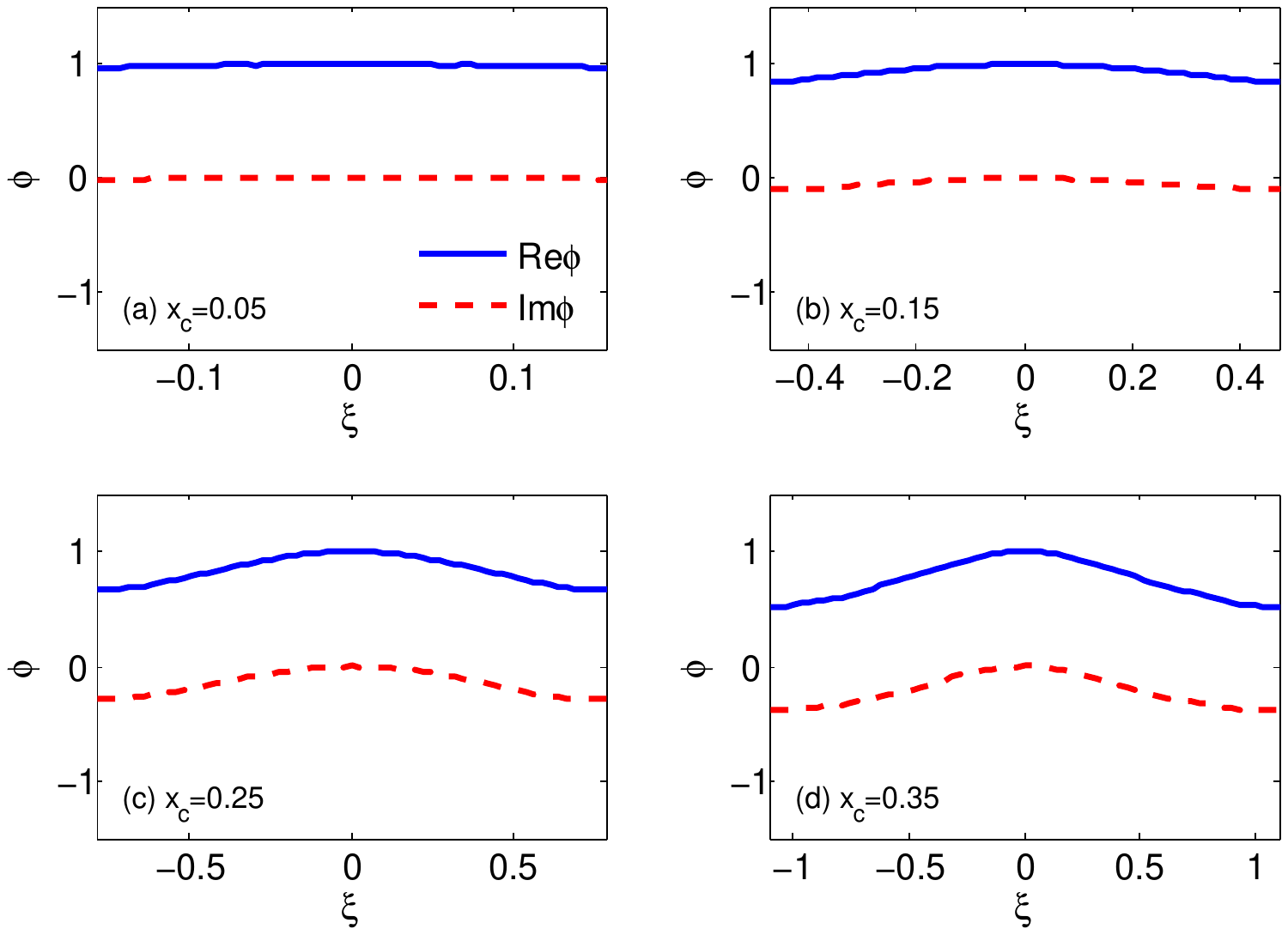}
\caption{Mode structures for $\kappa_n=5.0$, $\kappa_T=0.0$ at point
dipole with $x_c=0.05$, $0.15$, $0.25$ and $0.35$ respectively.}
\label{fig:phi_xi_point_xc}
\end{figure}

Figure \ref{fig:plt_w_k_dip} shows further comparison of the
simulation result with different type of bounce averages of the
dispersion relation. We can find that the results are mainly
affected by the bounce average of $\omega_D$, and the growth rate is
larger after bounce average which agree with the simulation result
in Fig.\ref{fig:bench_01}(d). The bounce average of $J_0$ affects
little to the real frequency and growth rate.
Fig.\ref{fig:phi_xi_point_k} shows the corresponding mode structures
for different $k_\perp\rho_i$, which shows that the mode structures
are not flat with $k_\perp\rho_i$ increasing and thus the
$k_\parallel=0$ assumption will be broken. This may explain the
larger deviation of the $\omega$ in Fig.\ref{fig:plt_w_k_dip} for
larger $k_\perp\rho_i$. Fig.\ref{fig:phi_xi_point_xc} shows the
corresponding mode structures for different $x_c$, where the
$k_\parallel=0$ assumption also not always holds.

The mode in scanning $x_c$ in Fig.\ref{fig:bench_01} only has slight
quantitative change of real frequency and growth rate. This tells us
that the physical feature of this mode does not depend on whether
the particles are trapped or passing. Thus, if we ignore the
influence of the mode structures, we can expect that the general
picture of the electrostatic drift modes in dipole configuration
(all are trapped particles) be similar to the one in Z-pinch (all
are passing particles) as studied in Ref.\cite{Ricci2006} and more
details in Ref.\cite{Xie2017a}, i.e., only two types of unstable
electrostatic drift modes in the system, one is mainly driven by
electron gradient and propagates in electron diamagnetic direction;
and another is mainly driven by ion gradient and propagates in ion
direction. However, when considering the mode structures, we find
some new physics not existed in Z-pinch configuration, i.e., high
order eigenstates of the mode can be more unstable than the ground
state in particular parameters.

The modes property in $(\kappa_n,\kappa_T)$ space with
$k_\perp\rho_i=0.5$ is shown in Fig.\ref{fig:scan_kap_2d}, which is
obtained from dispersion relation with matrix method as described in
Ref.\cite{Xie2017a} with only bounce averaged for $\omega_D$ not for
$J_0$. In this figure we can find two unstable modes exist, one is
in electron direction and another is in ion direction and the
unstable threshold for both $\kappa_n$ and $\kappa_T$ are around
$5.0$. The $\omega_r<0$ ion direction mode in
Fig.\ref{fig:scan_kap_2d}(a\&b) can be unstable for either larger
$\kappa_n$ or $\kappa_T$; whereas the $\omega_r>0$ electron
direction mode in Fig.\ref{fig:scan_kap_2d}(c\&d) is only unstable
when $\eta=\kappa_T/\kappa_n$ is large.


{\color{blue}We notice that this is opposite as the case in tokamak
community. In tokamak\cite{Garbet2006}, the ion (diamagnetic)
direction ion temperature gradient mode (ITG) is only unstable at
large $\eta$, whereas the electron direction trapped electron mode
(TEM) can be unstable with the $\eta=0$ density gradient driven
alone\footnote{The calculations of the entropy mode $\omega_r$ in
the Figs.1-4 in Ref.\cite{Xie2017a} are correct. However, we have
made mistake in the description in the last sentence of the first
paragraph of page 072106-5.}.}

\section{Parameters scan in Simulations}\label{sec:sim_result}

\begin{figure}[htbp]
\centering
\includegraphics[width=8cm]{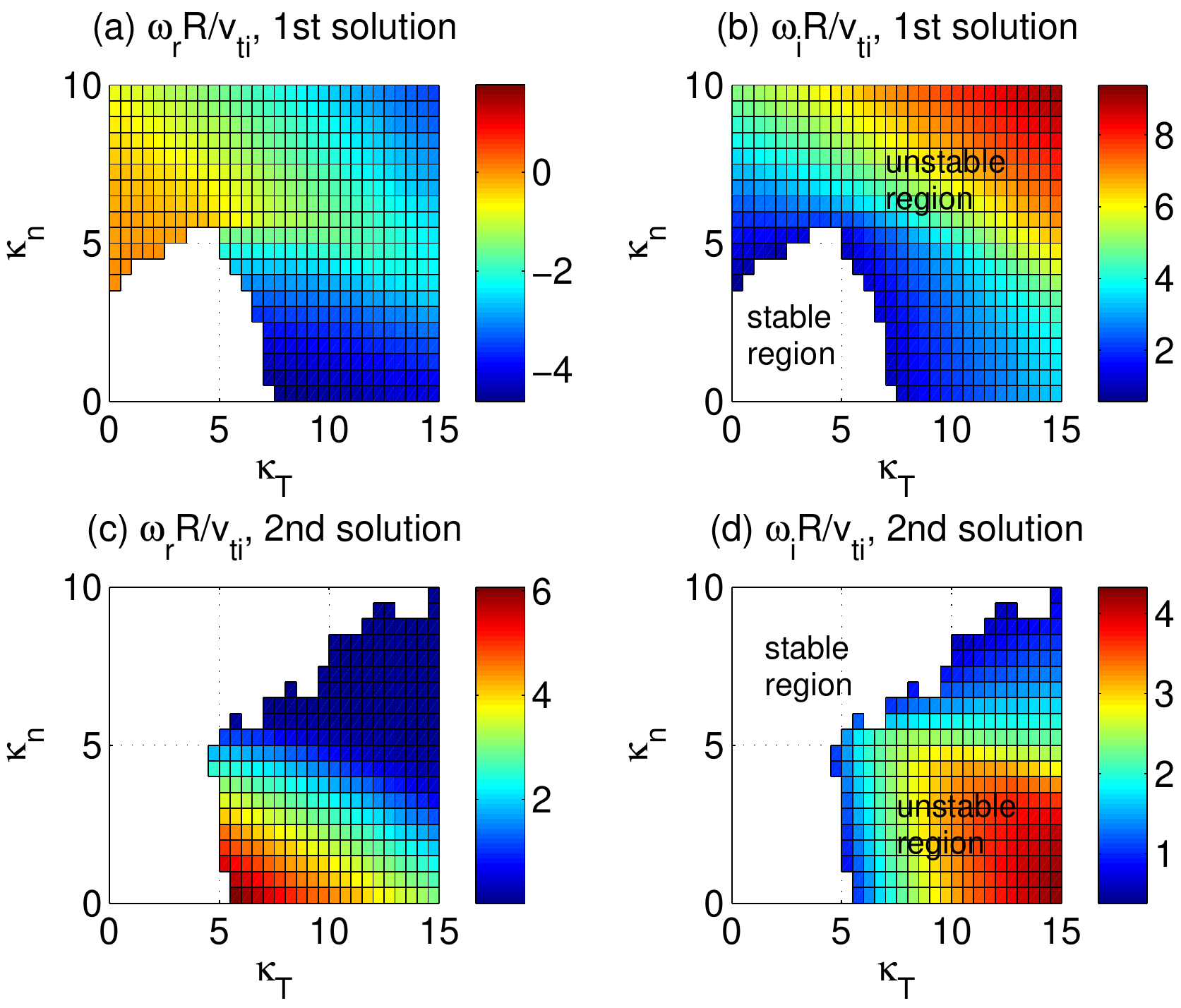}
\caption{Scan $(\kappa_n,\kappa_T)$ in point dipole using
dispersion relation, with $k_\perp\rho_i=0.5$. Two branches of
unstable mode exist: one in ion direction (a\&b) and another in
electron direction (c\&d).} \label{fig:scan_kap_2d}
\end{figure}

\begin{figure}[htbp]
\centering
\includegraphics[width=8cm]{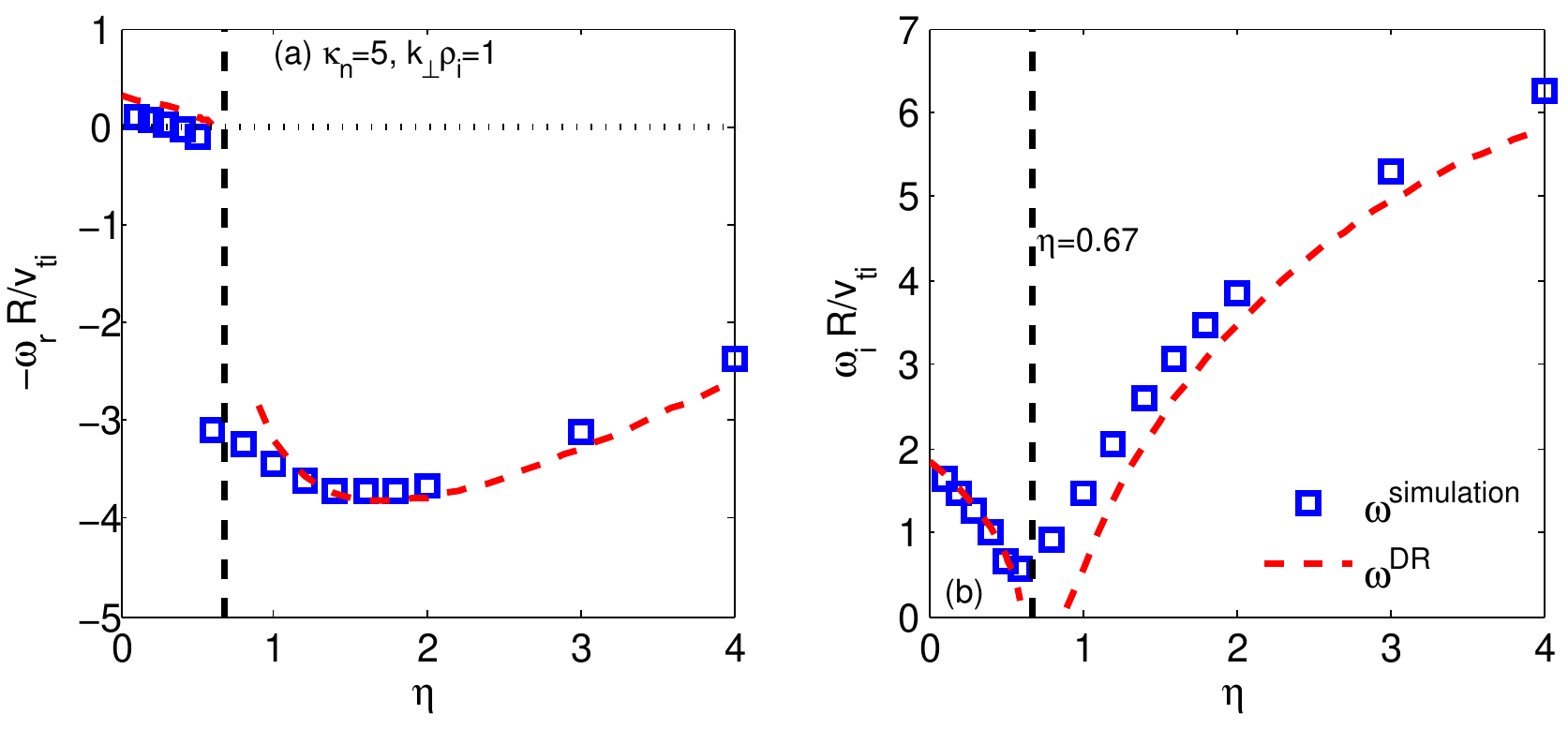}
\caption{Scan $\eta=\kappa_T/\kappa_n$ in point dipole and
comparison with dispersion relation solution.} \label{fig:scan_eta}
\end{figure}

\begin{figure}[htbp]
\centering
\includegraphics[width=8cm]{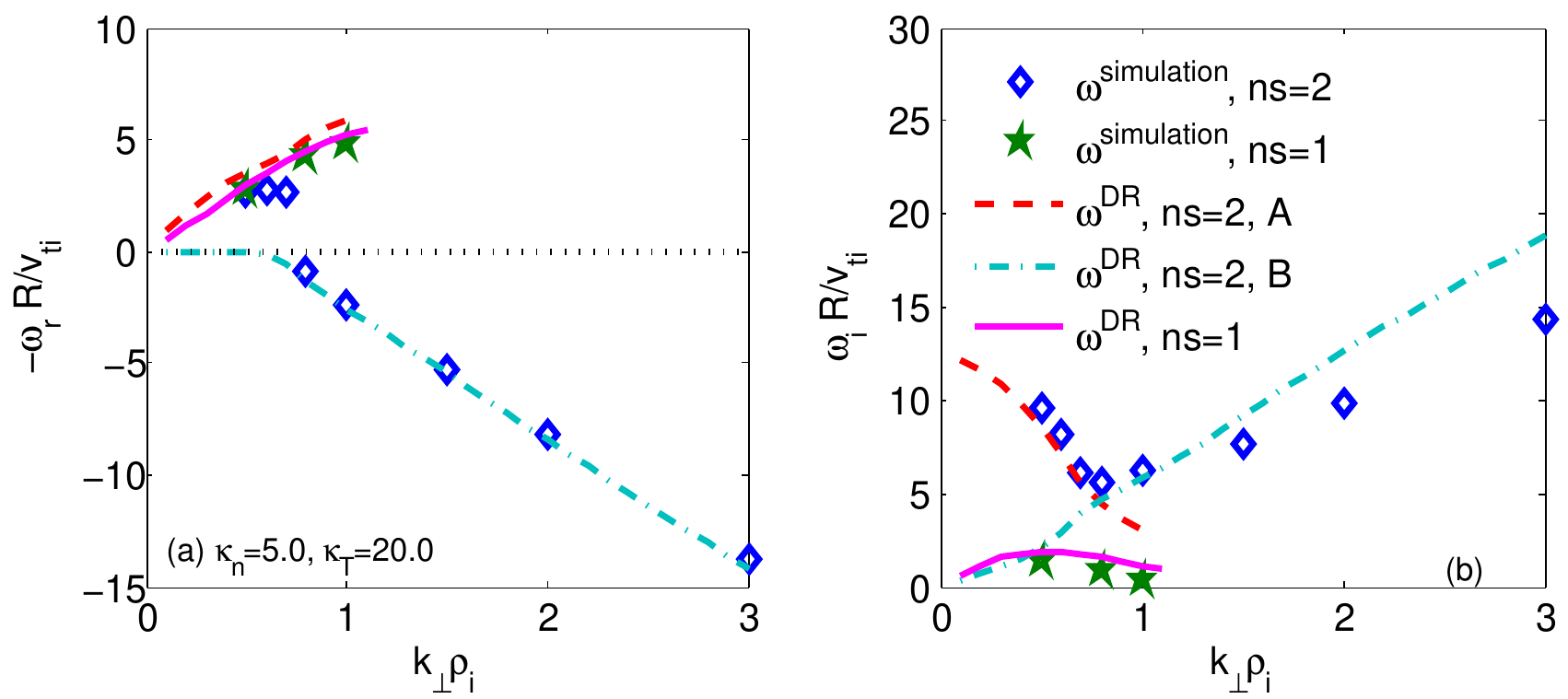}
\caption{Scan $k_\perp\rho_i$ in point dipole at $\eta=4$. With the
comparison of kinetic electron model ($ns=2$) and adiabatic electron
($ns=1$) model. } \label{fig:scan_ns_k}
\end{figure}

\begin{figure}[htbp]
\centering
\includegraphics[width=8cm]{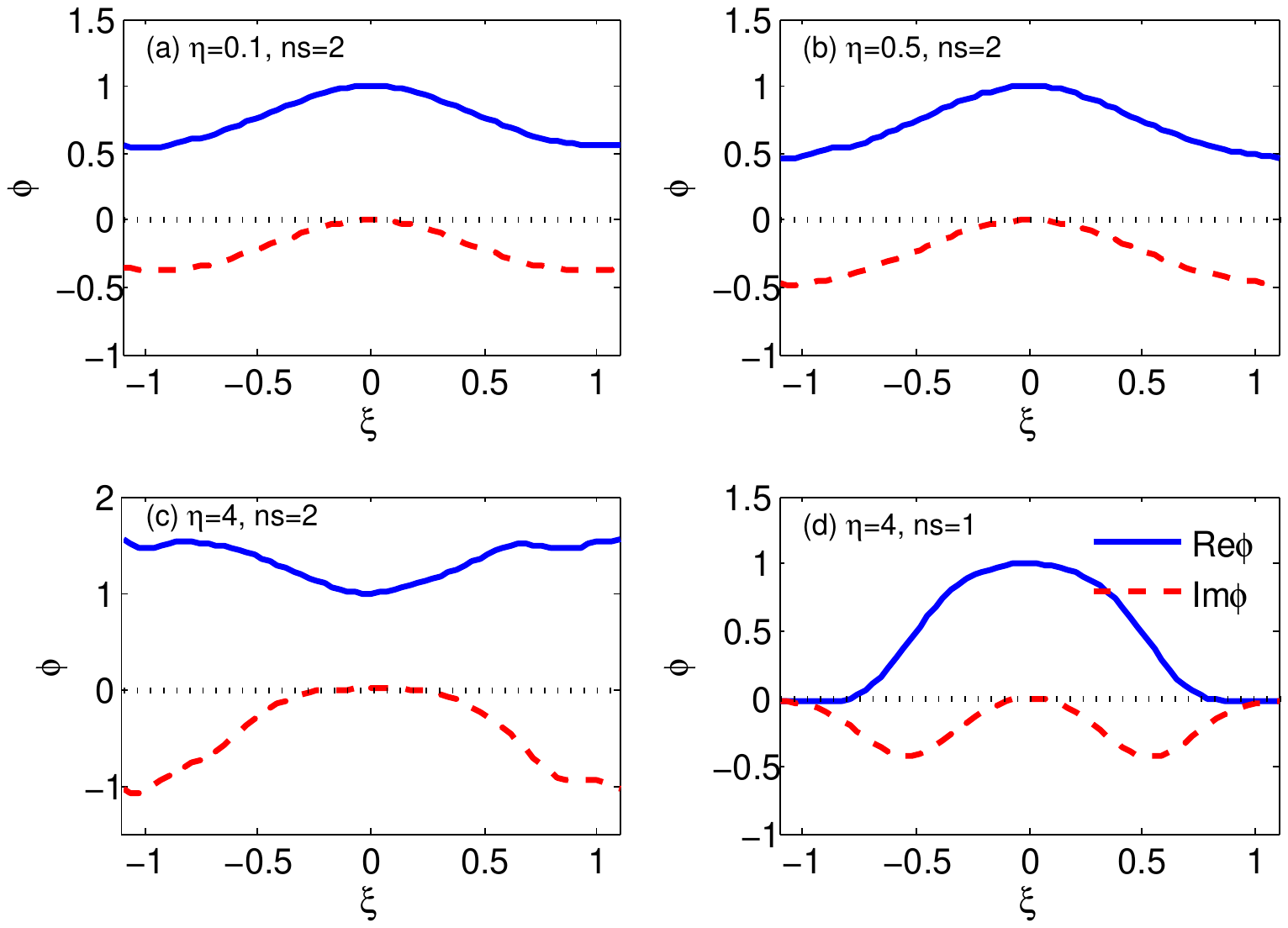}
\caption{Mode structures for $\kappa_n=5.0$, $k_\perp=1.0$ at point
dipole with $\eta=0.1$, $0.5$, $4.0$ with $ns=2$ and $\eta=4.0$ with
$ns=1$ respectively.} \label{fig:phi_xi_point_eta}
\end{figure}

In this section, we study details of the linear electrostatic drift
modes in the dipole configuration. In our simulations, we find the
qualitative features of these electrostatic drift modes are similar
between point dipole and ring dipole. And the differences are only
quantitatively. Thus, we mainly focus on point dipole case. Firstly
we scan $\eta$ to confirm that two types of mode exist in the
system. Figure \ref{fig:scan_eta} shows the results with
$\kappa_n=5.0$, $k_\perp\rho_i=1.0$, $x_c=0.35$, $n_p=5\times10^5$
$n_g=64$, {\color{blue}$dt=0.0002$ and $n_t=3\times10^5$}. We can
see that indeed there exists a transition from ion mode with
negative frequency to electron mode with positive frequency at
around $\eta\sim\eta_c=2/3$. In Fig.\ref{fig:scan_kap_2d}, we know
that both electron and ion modes can be unstable at large $\eta$.
The results of scan $k_\perp\rho_i$ for a typical large $\eta$ are
shown in Fig.\ref{fig:scan_ns_k}, where we find that the ion mode is
dominant at small $k_\perp\rho_i$ and the electron mode becomes
dominant at larger $k_\perp\rho_i$. The dispersion relation can
still predict the qualitative features of these two modes. In
Fig.\ref{fig:scan_ns_k}, we have also shown the adiabatic electron
model ($ns=1$) results, which is in the ion direction. Although the
real frequency can roughly agree with the kinetic electron model
($ns=2$) results, the behavior of the growth rate is much different
between these two models. This tells us that the electrostatic drift
modes in dipole configuration can not be described by adiabatic
electron model. Figure \ref{fig:phi_xi_point_eta} shows the
corresponding mode structures by scanning $\eta$ and the structure
of $ns=1$ mode is much center peaked and not flat as the $ns=2$
mode. Here and after, the dispersion relation results in point
dipole are all obtained with bounce averaged only for $\omega_D$ and
not for $J_0$. In the latter part, we will mainly focus on the
$\eta<\eta_c$ mode.


\begin{figure}[htbp]
\centering
\includegraphics[width=8cm]{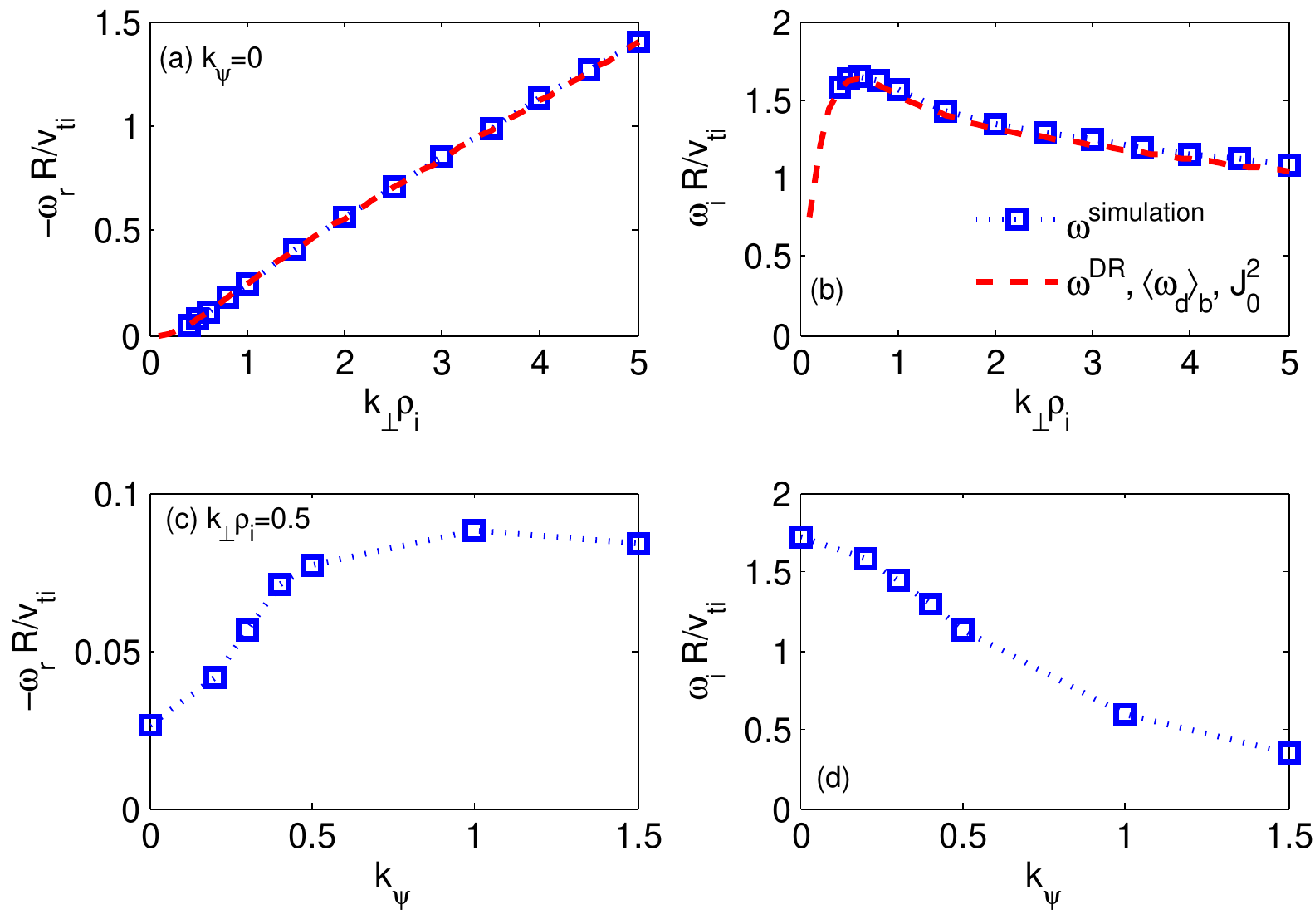}
\caption{Scan of $k_\perp\rho_i$ and $k_\psi$, with $\kappa_n=5.0$
and $\kappa_T=1.0$ in point dipole.} \label{fig:scan_kzeta_kpsi}
\end{figure}

\begin{figure}[htbp]
\centering
\includegraphics[width=8cm]{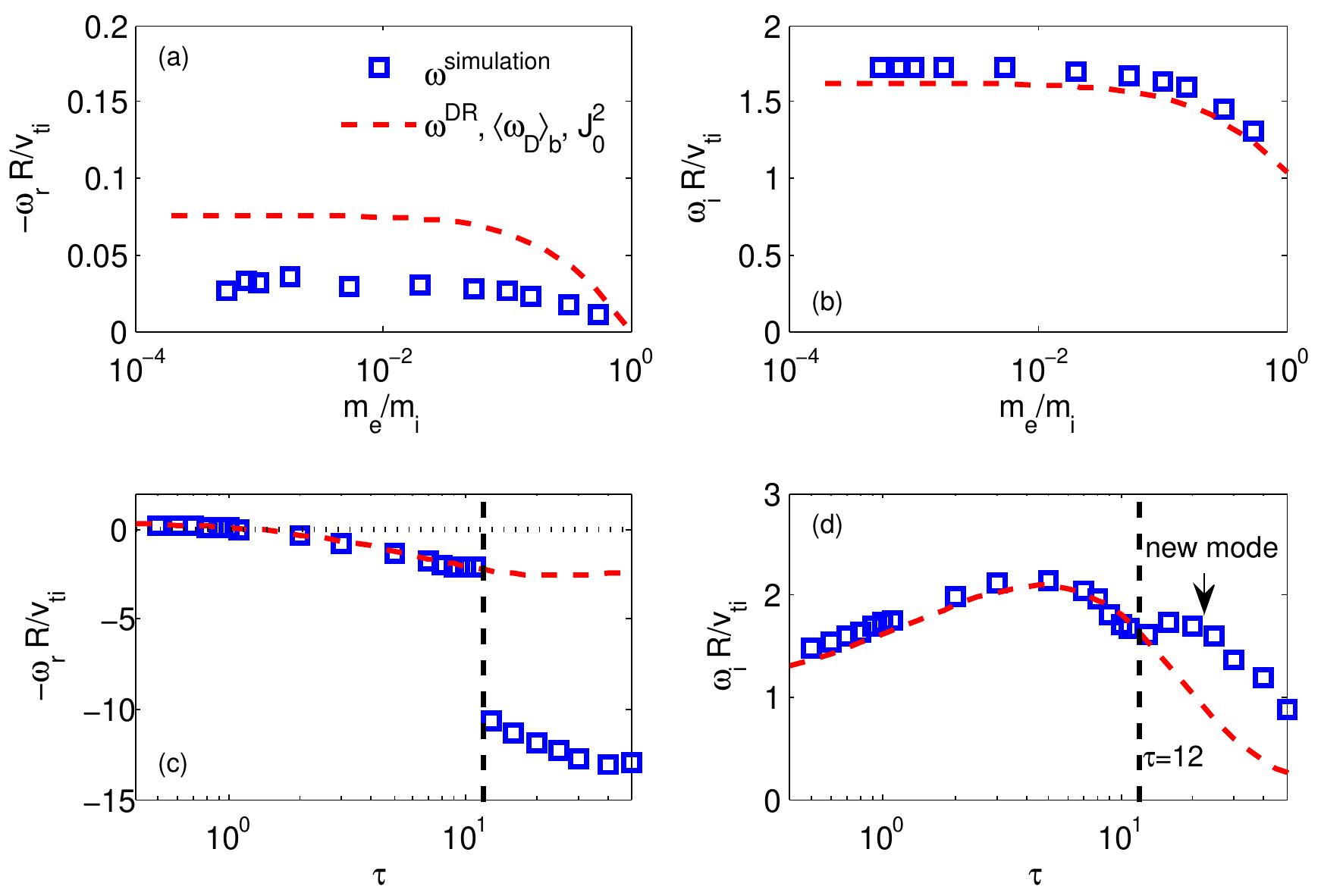}
\caption{Scan of $m_e/m_i$ and $\tau$, with $\kappa_n=5.0$,
$\kappa_T=1.0$ and $k_\perp\rho_i=0.5$ in point dipole.}
\label{fig:scan_m_tau}
\end{figure}

\begin{figure}[htbp]
\centering
\includegraphics[width=8cm]{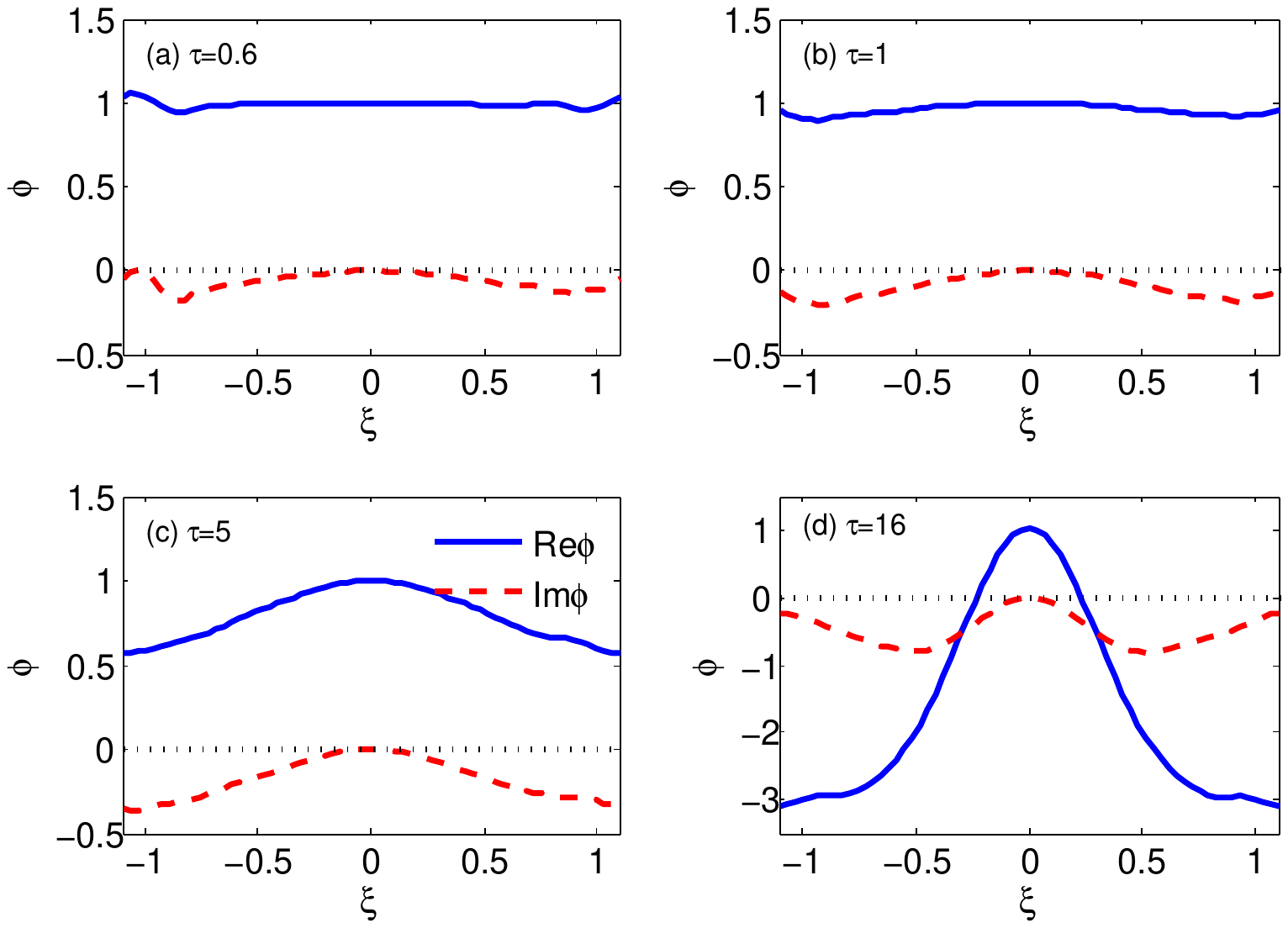}
\caption{Mode structures for $\kappa_n=5.0$, $\kappa_T=1.0$ and
$k_\perp\rho_i=0.5$ in point dipole with $\tau=0.5$, $1$, $5$ and
$16$ respectively.} \label{fig:phi_xi_point_tau}
\end{figure}

\begin{figure}[htbp]
\centering
\includegraphics[width=8cm]{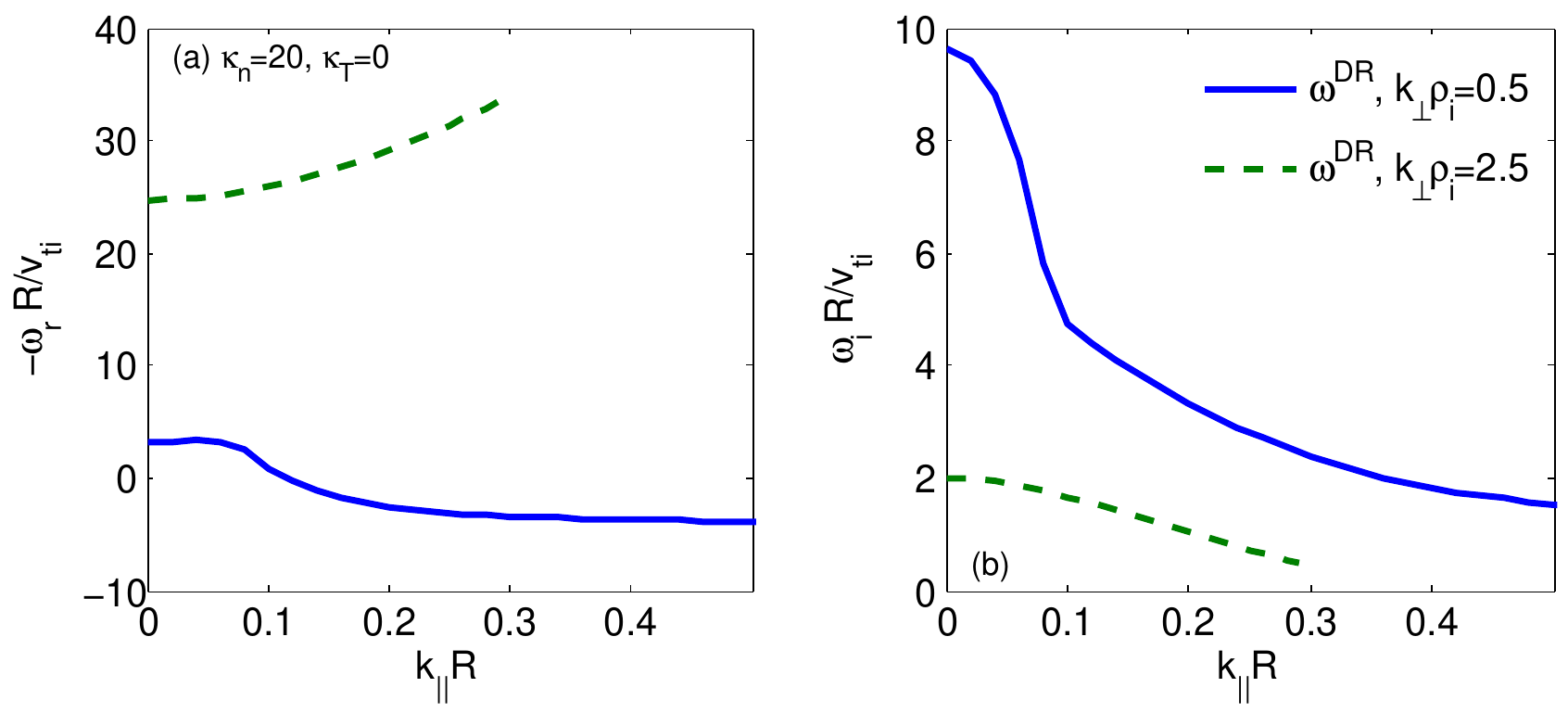}
\caption{Scan of $k_\parallel R$ using the dispersion relation
Eq.(\ref{eq:entropy_DR}), with $\kappa_n=20$ and $\kappa_T=0$ in
point dipole, for $k_\perp\rho_i=0.5$ and $k_\perp\rho_i=2.5$.}
\label{fig:scan_kz}
\end{figure}

Figure \ref{fig:scan_kzeta_kpsi} shows the scan of $k_\zeta$ and
$k_\psi$. The result for $k_\zeta$ scanning agrees with the typical
entropy mode feature in Ref.\cite{Ricci2006,Xie2017a}, and
especially for large $k_\perp\rho_i$ the mode is still unstable. The
most unstable solution exists at $k_\perp\rho_i\simeq1.0$. The
$k_\psi$ scan does not bring any qualitative difference but play a
role in increasing $k_\perp$ [see Eq.(\ref{eq:kperp})]. Figure
\ref{fig:scan_m_tau}(a\&b) shows the scan of $m_e/m_i$. We find that
at small $m_e/m_i<0.01$ the simulation results of $\omega_r$ and
$\gamma$ change little. Whereas the scan of $\tau$ in
Fig.\ref{fig:scan_m_tau}(c\&d) show something interesting, i.e., a
new mode become unstable at $\tau>12$. And around $\tau=12$, two
modes compete in the simulation. The real frequency is very small at
$\tau\sim1$, where a smooth sign change from ion direction to
electron direction is observed. By scan $\tau$ in dispersion
relation solver, we only find one unstable mode, i.e., this new mode
may not exist in the zero-dimensional model and must have something
to do with the mode structure. We find indeed that these two modes
have different mode structures. As shown in
Fig.\ref{fig:phi_xi_point_tau}, at small $\tau$ ($\tau=0.6$ and
$1.0$) the mode structure of $\phi(\xi)$ is flat; at mediate $\tau$
($\tau=5.0$) $\phi(\xi)$ is much similar as the one in
Fig.\ref{fig:phi_xi_point_k}(d) and
Fig.\ref{fig:phi_xi_point_xc}(d); for large $\tau$
($\tau=16>\tau_c\simeq12$) the mode structure changes more, e.g.,
both $Re\phi(\xi)>0$ and $Re\phi(\xi)<0$ exist. This jump also
exists in ring dipole case. This new mode may be the high order
eigenstate of the original ground state mode. We will show that
series of high order eigenstates indeed exist in the system and can
be the most unstable one. As shown in Fig.\ref{fig:scan_kz}, we scan
the zero-dimensional dispersion relation Eq.(\ref{eq:entropy_DR})
with $k_\parallel\neq0$ and find that only one unstable mode exists
and the $k_\parallel$ only brings damping effect because of electron
Landau damping. This implies that this one-dimensional new mode in
Fig.\ref{fig:scan_m_tau}(c\&d) can not be predicted by
zero-dimensional dispersion relation.

\begin{figure}[htbp]
\centering
\includegraphics[width=8cm]{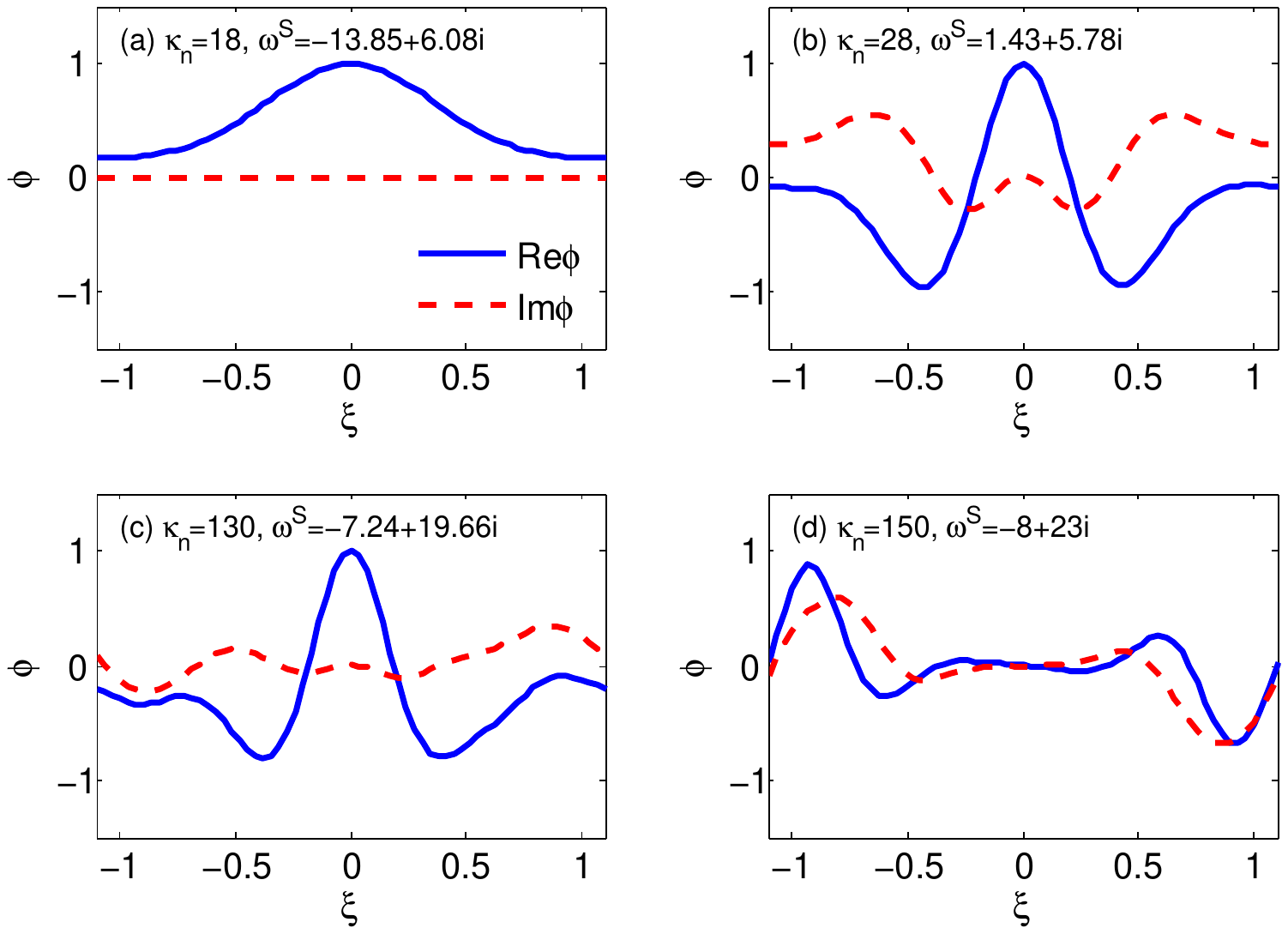}
\caption{High order eigenstates mode exist at strong gradient with
$\kappa_T=0.0$, $k_\perp=2.5$ and $\kappa_n=18$, $28$, $130$ and
$150$ respectively in point dipole configuration.}
\label{fig:phi_xi_point_kapn}
\end{figure}

\begin{figure}
 \centering
  \includegraphics[width=8.0cm]{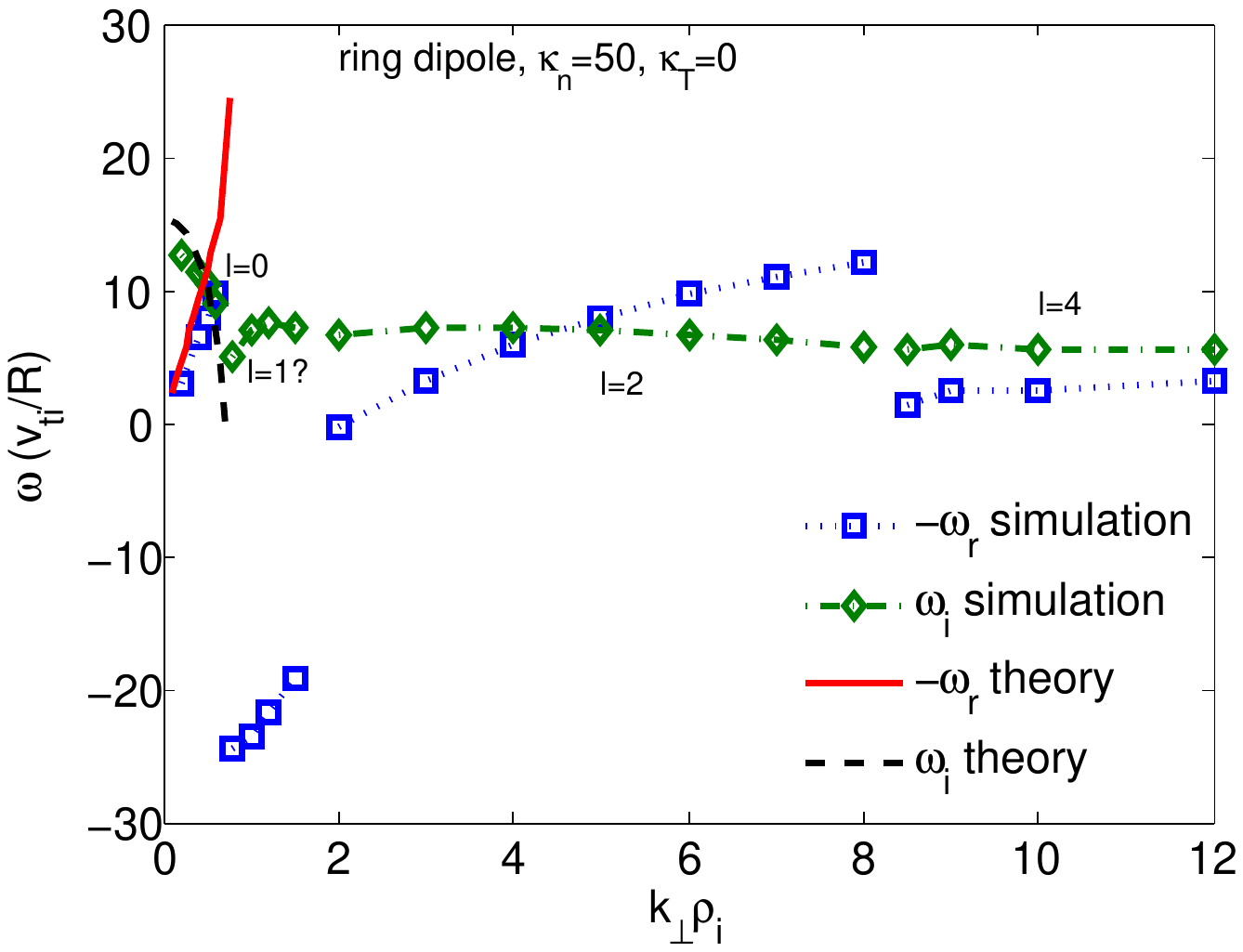}\\
  \caption{Scanning $\omega$ vs. $k_\perp\rho_i$ for $\kappa_n=50$,
  $\kappa_T=0$ in ring dipole configuration. High order modes are most unstable for large $k_\perp\rho_i$.
  }\label{fig:w_kt_l}
\end{figure}

\begin{figure}
 \centering
  \includegraphics[width=8.0cm]{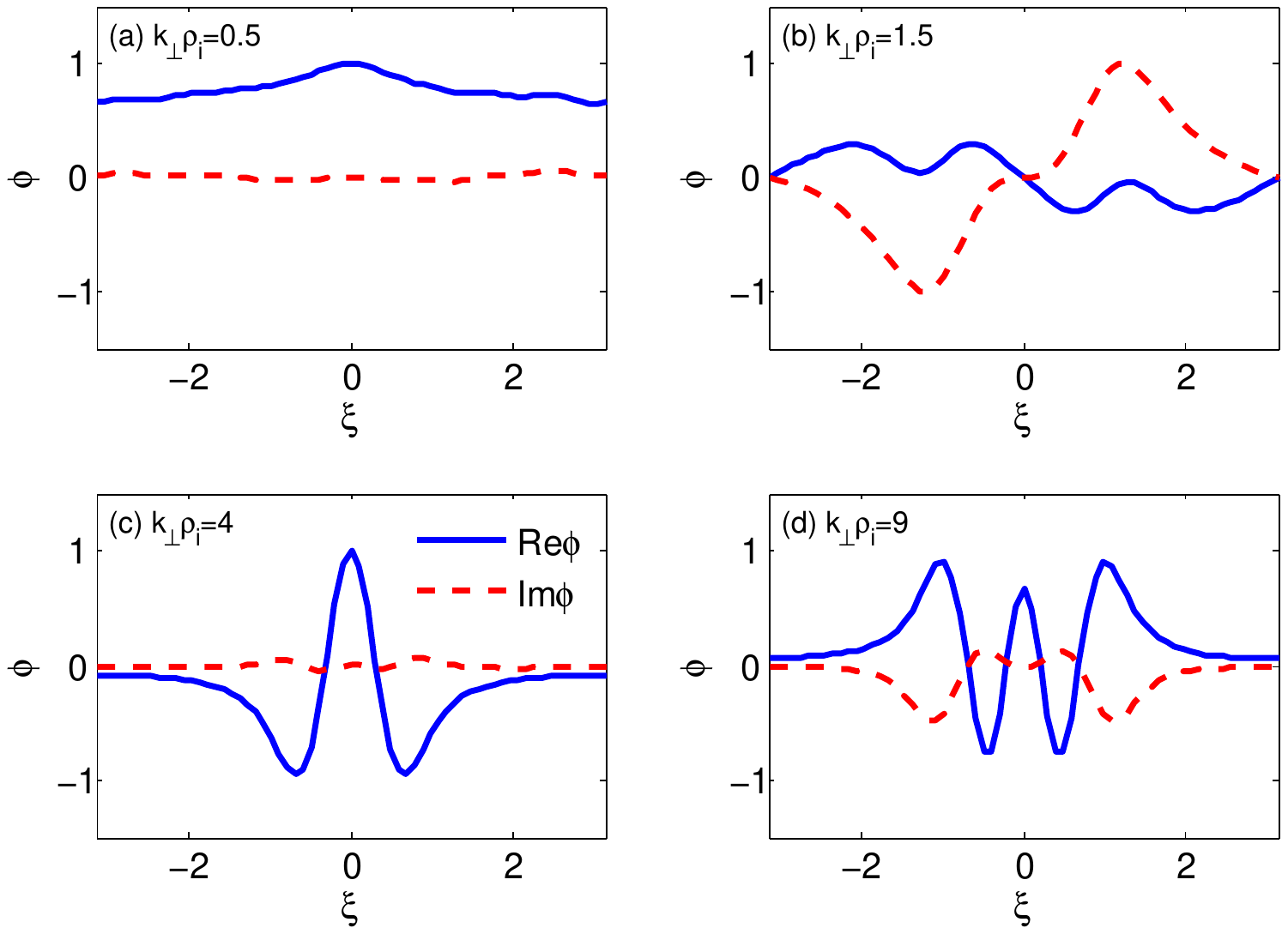}\\
  \caption{Corresponding mode structures vs. $k_\perp\rho_i$ for $\kappa_n=50$,
  $\kappa_T=0$ in ring dipole configuration.
  }\label{fig:phi_xi_ring_k}
\end{figure}

By increasing density gradient $\kappa_n$ in
Fig.\ref{fig:phi_xi_point_kapn}, we find the most unstable mode in
the system can have very different mode structures.
Fig.\ref{fig:phi_xi_point_kapn}(a) shows the ground mode;
Fig.\ref{fig:phi_xi_point_kapn}(b\&c) show high-order even mode; and
Fig.\ref{fig:phi_xi_point_kapn}(d) shows high-order odd mode.
Considering that the result may be affected by $x_c$ in the point
dipole simulation, we study this feature in detail using ring dipole
configuration with $x_c=1.0$, i.e., to remove the effects of the
boundary condition. Figure \ref{fig:w_kt_l} shows a scan of
$k_\perp\rho_i$ in the ring dipole configuration with $\kappa_n=50$.
We see that the dispersion relation only predicts an unstable mode
at $k_\perp\rho_i<1.5$, which qualitatively agrees with the $l=0$
mode in simulation. The slab dispersion relation is used to obtain
$\omega^{\rm theory}$, but we have used the curvature drift
frequency at $\xi=0$, i.e., $\omega_d^{\rm ring}=g(0)\omega_d^{\rm
Z-pinch}$ with $g(0)=1.31$. However, new unstable modes appear by
increasing $k_\perp\rho_i$ in the simulation, and the corresponding
mode structures are shown in Fig.\ref{fig:phi_xi_ring_k}. Due to
steep gradient, this mode is `interchange-like' \cite{Xie2017a}, as
$\omega_i\to\gamma_0\neq0$ for $k_\perp\rho_i\to0$. The eigenstate
label $l$ here is roughly to fit the mode structure with
$\phi(\xi)=H_l(\xi)e^{-\xi^2/2}$, where $H_l$ is $l$-th Hermite
polynomials, with $l=0,1,2,\cdots$. That is, under strong gradient
or large $\tau$, the most unstable mode in the system can be on
non-ground state and the assumption $k_\parallel=0$ is not valid any
more, which is not predicted by conventional understandings (cf.
Refs.\cite{Kobayashi2009,Kesner2002}). These high order eigenstates
have been predicted in tokamak edge steep gradient parameters
recently\cite{Xie2015} where the physical explanation of it is the
change of quantum potential well, and which can also change the
nonlinear transport feature \cite{Xie2017}. It is also interesting
that the even mode and odd mode have opposite propagation directions
in Fig.\ref{fig:phi_xi_ring_k}. The physical reason for why these
high order modes can be most unstable is yet to study. The
drift-bounce resonance $\omega-p\omega_b-q\omega_D=0$ with
$p,q=0,\pm1,\pm2,\cdots$ in Refs.\cite{Dettrick2003,Zhu2014} is one
possibility. However, we did not find clear resonant structures in
$g_{i,e}(E,\lambda)$ velocity space as in Ref.\cite{Zhu2014} yet.
The velocity space resonance in both
Refs.\cite{Dettrick2003,Zhu2014} is between energetic ions and
electromagnetic Alfv\'en mode, whereas our simulation includes only
background ions and electrostatic perturbations. It is also not
clear yet how this high order modes affect the nonlinear physics in
dipole plasma.

\section{Summary and Conclusion}\label{sec:summ}

In this work, we have developed a 1D linear local collisionless
gyrokinetic $\delta f$ PIC code to study the electrostatic drift
modes in a point and ring dipole plasmas. With assumption of
$k_\parallel=0$, the corresponding 0D bounce averaged dispersion
relation is also derived and solved for benchmark. We find the
general feature of the electrostatic drift modes in dipole
configuration is similar to the one in Z-pinch configuration, i.e.,
two unstable modes exist mainly at either small $\eta$ or large
$\eta$ with critical $\eta_c\simeq2/3$. However, there still exists
much difference between dipole and Z-pinch configurations. In
Z-pinch, $k_\parallel\simeq0$ is a good approximation; whereas for
point and ring dipole, $k_\parallel\simeq0$ is only valid at very
narrow parameter space, e.g., only for small $k_\perp\rho_i$. Some
new unstable modes with non-flat mode structure are found in dipole
configuration at large $\tau$ and $\kappa_n$. With a smooth change
of the temperature ratio $\tau=T_e/T_i$ from 0.5 to 60, we find a
jump in real frequency and a turning point in growth rate at around
$\tau =\tau_c\simeq12$ which is caused by the competition between
two different modes. More clearly, we have demonstrated that the
most unstable mode is at high order eigenstates with either odd
parity or even parity at large $\kappa_n$, as has been predicted
\cite{Xie2015} in tokamak configuration.

We also notice that one major difference between dipole
configuration and tokamak is the magnetic shear $s=0$. For $s\neq0$
as in tokamak, the parallel boundary condition in the simulation
model should be modified\cite{Beer1995,Chen2003} and the physics can
be different. And the ITG and TEM in tokamak usually have
$k_\parallel\neq0$. What we think is interesting is that although
almost all the particles are trapped in the dipole configuration, we
have not found new mode can be called as trapped electron
mode\cite{Coppi1974} as in tokamak. The essential unstable
electrostatic drift modes in dipole configuration is similar to the
one in Z-pinch configuration where all particles are passing
particles.

In summary, we have given a comprehensive linear study of the 1D
electrostatic drift modes in a point and ring dipole plasma. This
helps us to understand the basic linear behaviors of the unstable
modes in the system and can provide a starting point for further
nonlinear study. Contrast to previous studies \cite{Kesner2000,
Simakov2001}, this work is valid for all the gyrokinetic orderings,
without further approximations. Contrast to previous Z-pinch and
ring dipole
studies\cite{Ricci2006,Kobayashi2009,Kobayashi2010,Xie2017a}, this
work demonstrates the importance of the parallel mode structures and
finds that the high order eigenstates can be important at some
parameters.

\acknowledgments The authors would like to thank M. Mauel for
valuable discussions and B. Rogers for suggestions to include ring
dipole configuration. Communications with J. Kesner, Y. Y. Li, P.
Porazik, Z. Lin, A. Bierwage, L. J. Zheng and L. Chen are also
acknowledged. The work was supported by Natural Science Foundation
of China under Grant No. 11675007 and the China Postdoctoral Science
Foundation No. 2016M590008.

\appendix

\section{Current loop/ring dipole magnetic field}\label{sec:apd_ring_eq}

We consider the dipole magnetic field produced by single current
loop \cite{Jackson1999} with radius $a$. In cylinder coordinate
$(\rho,\phi,z)$, for current density
\begin{equation}\label{eq:mirror_current}
{\bf{J}} = I\delta \left( {z'} \right)\delta \left( {\rho ' - a}
\right)\left( { - \sin \phi ',\cos \phi ',0} \right),
\end{equation}
we have the magnetic vector potential
\begin{equation}\label{eq:mirror_A}
{\bf{A}} = {{{\mu _0}} \over {4\pi }}\int {{{{\bf{J}}\left(
{{\bf{r'}}} \right)} \over {\left| {{\bf{r}} - {\bf{r'}}}
\right|}}{d^3}{\bf{r'}}},
\end{equation}
where $a$ is the radius of the current loop.

Due to symmetry, magnetic vector has only $\phi$ component, and
\begin{equation}\label{eq:mirror_A_phi}
{A_\phi } = {{{\mu _0}I} \over {2\pi }}\sqrt {{a \over \rho }}
\left[ {{{\left( {{k^2} - 2} \right)K\left( {{k^2}} \right) +
2E\left( {{k^2}} \right)} \over {2k}}} \right],
\end{equation}
where $K$ and $E$ are the first and second kind of elliptic
functions with
\begin{equation}\label{eq:mirror_EK_k2}
    k^2=\frac{4a\rho}{(a+\rho)^2+z^2},~~\rho^2=x^2+y^2.
\end{equation}
Hence, we obtain the magnetic field
\begin{equation}\label{eq:mirror_A_phi}
{{\bf{B}}_{{\rm{singlecoil}}}} = \nabla  \times {\bf{A}} =  -
{{\partial {A_\phi }} \over {\partial z}}\hat \rho  + {1 \over \rho
}{{\partial \left( {\rho {A_\phi }} \right)} \over {\partial \rho
}}\hat z.
\end{equation}

Using
\begin{equation}\label{eq:mirror_dEK}
\left\{ \begin{aligned}
  K'\left( x \right) &= {1 \over {2x\left( {1 - x} \right)}}E\left( x \right) - {1 \over {2x}}K\left( x \right),
  \\
  E'\left( x \right) &= {1 \over {2x}}E\left( x \right) - {1 \over {2x}}K\left( x \right), \\ \end{aligned}  \right.
\end{equation}
we obtain{\small
\begin{equation}\label{eq:mirror_B3}
\left\{ \begin{aligned}
  {B_\rho } &= {{{\mu _0}I} \over {2\pi }}{z \over {\rho \sqrt {{{\left( {a + \rho } \right)}^2}
  + {z^2}} }}\left[ {{{{a^2} + {\rho ^2} + {z^2}} \over {{{\left( {a - \rho } \right)}^2}
  + {z^2}}}E\left( {{k^2}} \right) - K\left( {{k^2}} \right)} \right],
  \\
  {B_z} &= {{{\mu _0}I} \over {2\pi }}{1 \over {\sqrt {{{\left( {a + \rho } \right)}^2}
  + {z^2}} }}\left[ {{{{a^2} - {\rho ^2} - {z^2}} \over {{{\left( {a - \rho } \right)}^2}
  + {z^2}}}E\left( {{k^2}} \right) + K\left( {{k^2}} \right)} \right],
  \\
  {B_\phi } &= 0.  \end{aligned}  \right.
\end{equation}}

And the magnetic flux can be calculated as $\psi=\oint {\bm B} \cdot
d{\bm S}=\oint (\nabla\times {\bm A})  \cdot d{\bm S}=\oint {\bm A}
\cdot d{\bm l}=2\pi R A_\phi$, which would be used to determine the
flux surface. For practise usage, we calculate the field line
functions numerically and use interpolation basing on the above
formula.

\section{Point dipole operators}\label{sec:apd_point_op}
Comparing to the ring dipole configuration equations in
Sec.\ref{sec:dip_op}, the benefit of using ideal point dipole
configuration is that all the operators we used in the numerical
code can be obtained analytically. We can have the following
orthogonal flux coordinate $(\psi,\chi,\zeta)$ for ideal point
dipole configuration{\small
\begin{equation}\label{eq:flux_coor1}
\left\{ \begin{split}
      \psi & = \frac{M\sin^2\theta}{r},\\
      \chi & = \frac{M\cos\theta}{r^2},\\
      \zeta & = \phi,
    \end{split}
    \right.~~\left\{ \begin{split}
      {\bm e}^\psi & =\nabla\psi= \frac{M\sin\theta}{r^2}(-\sin\theta{\bm \hat e}_r+2\cos\theta{\bm \hat e}_\theta),\\
      {\bm e}^\chi & =\nabla\chi= -\frac{M}{r^3}(2\cos\theta{\bm \hat e}_r+\sin\theta{\bm \hat e}_\theta),\\
      {\bm e}^\zeta & =\nabla\zeta= \frac{1}{r\sin\theta}{\bm \hat e}_\phi,
    \end{split}
    \right.
\end{equation}}
where $M$ is the magnetic moment of the dipole, and
$(r,\theta,\phi)= (radial, polar, azimuthal)$ is spherical
coordinates with ${\bm \hat e}_r$, ${\bm \hat e}_\theta$ and ${\bm
\hat e}_\phi$ the unit vector in each directions. Note that ${\bm
B}=\nabla\phi\times\nabla\psi=\nabla\chi$ and $B=\sqrt{{\bm
B}\cdot{\bm B}}=(M/r^3)\sqrt{1+3\cos^2\theta}$.

For completeness, besides the contravariant vector ${\bm
e}^{\alpha}=\nabla \alpha$, we also list here the covariant vector
${\bm e}_{\alpha}=\partial_\alpha {\bm r}$
\begin{equation}
\left\{ \begin{split}
      {\bm e}_\psi & =\partial_\psi{\bm r}= \frac{r^2(-\sin\theta{\bm \hat e}_r+2\cos\theta{\bm \hat e}_\theta)}{M\sin\theta(1+3\cos^2\theta)},\\
      {\bm e}_\chi & =\partial_\chi{\bm r}= -\frac{r^3(2\cos\theta{\bm \hat e}_r+\sin\theta{\bm \hat e}_\theta)}{M(1+3\cos^2\theta)},\\
      {\bm e}_\zeta & =\partial_\zeta{\bm r}= r\sin\theta{\bm
      \hat e}_\phi,
    \end{split}
    \right.
\end{equation}
with also ${\bm e}^{\alpha}\times{\bm
e}^{\beta}=\mathcal{J}^{-1}{\bm e}_{\gamma}$ and ${\bm
e}^{\alpha}\cdot{\bm e}_{\beta}=\delta^\alpha_\beta$. And it is
readily to obtain the contravariant metric tensor
$g^{\alpha\beta}\equiv\nabla\alpha\cdot\nabla\beta$, the covariant
metric tensor $g_{\alpha\beta}\equiv\partial_\alpha{\bm
r}\cdot\partial_\beta{\bm r}$, and the Jacobian
$\mathcal{J}\equiv{\bm e}_{\alpha}\cdot{\bm e}_{\beta}\times{\bm
e}_{\gamma}=1/\sqrt{\det(g^{\alpha\beta})}=\sqrt{\det(g_{\alpha\beta})}=\frac{r^6}{M^2(1+3\cos^2\theta)}$.
Note also $g_{\alpha\beta}g^{\beta\gamma}=\delta_\alpha^\gamma$,
${\bm e}_\alpha=g_{\alpha\beta}{\bm e}^\beta$, ${\bm
e}^\alpha=g^{\alpha\beta}{\bm e}_\beta$, ${\bm e}_\alpha\times{\bm
e}_\beta=\mathcal{J}\varepsilon_{\alpha\beta\gamma}{\bm e}^\gamma$
and ${\bm e}^\alpha\times{\bm
e}^\beta=\mathcal{J}^{-1}\epsilon^{\alpha\beta\gamma}{\bm
e}_\gamma$. Considering covariant ${\bm B}=B_\alpha {\bm e}^\alpha$
($B_\alpha ={\bm B}\cdot{\bm e}^\alpha$) and contravariant ${\bm
B}=B^\alpha {\bm e}_\alpha$ ($B^\alpha ={\bm B}\cdot{\bm
e}_\alpha$), we can have ideal dipole field ${\bm B}={\bm
e}^\chi=\mathcal{J}^{-1}{\bm e}_\chi$, which is true because ${\bm
B}=\nabla\psi\times\nabla\zeta$ and
$\nabla\psi\times\nabla\zeta=\mathcal{J}^{-1}\partial_\chi {\bm r}$,
and we obtain
\begin{equation}\label{eq:flux_B}
\left\{ \begin{split}
      B^\chi & =\mathcal{J}^{-1}=\frac{M^2(1+3\cos^2\theta)}{r^6},\\
      B^\psi & =0,\\
      B^\zeta & =0,
    \end{split}
    \right.~~~~~~
\left\{ \begin{split}
      B_\chi & =1,\\
      B_\psi & =0,\\
      B_\zeta & =0,
    \end{split}
    \right.
\end{equation}
with magnetic vector ${\bm A}=-\psi\nabla\zeta$, i.e.,
$\nabla\times{\bm
A}=-\nabla\times\psi\nabla\zeta=\nabla\zeta\times\nabla\psi={\bm
B}$.

Along a field line $l$, $\psi$ and $\zeta$ do not change, and we can
still obtain $dl=r_0\kappa(\xi) d\xi$, where  $\xi=\pi/2-\theta$,
$\kappa=\cos\xi(1+3\sin^2\xi)^{1/2}$ and $r_0$ is the distance from
the flux surface to the origin at the equator $\xi=0$. Note that
here $\xi\in(-\pi/2,\pi/2)$, which differs from the ring dipole case
$\xi\in[-\pi,\pi]$. We can readily obtain
$f(\xi)=B/B_0=\frac{\sqrt{1+3\sin^2\xi}}{\cos^6\xi}$, and
\begin{eqnarray}\label{eq:kperp}
    k_\perp^2&=&k_{\psi}^2 \nabla \psi\cdot \nabla \psi+k_{\zeta }^2 \nabla \zeta\cdot \nabla
    \zeta\\\nonumber
    &=&\frac{1}{r_0^2\cos^6\xi}\Big[k_{\zeta
    }^2+k_{\psi}^2{B_0^2r_0^4}(1+3\sin^2\xi)\Big],
\end{eqnarray}
where we have used Eq.(\ref{eq:flux_coor1}), $r=r_0\sin^2\theta$ and
$B_0=M/r_0^3$. Here, we have $\rho_{s}
=\frac{v\sqrt{\lambda}}{\Omega_{s0}}\Big|\frac{\cos^3\xi}{(1+3\sin^2\xi)^{1/4}}\Big|$
and consider $k_{\psi}\neq0$, and thus
\begin{equation*}
  k_\perp\rho_{s} = \frac{v\sqrt{\lambda}}{r_0\Omega_{s0}}\sqrt{k_{\zeta
    }^2+k_{\psi}^2{B_0^2r_0^4}z}\frac{1}{z^{1/4}},
\end{equation*}
\begin{equation*}
  \partial_\xi( k_\perp\rho_s) = \frac{v\sqrt{\lambda}}{r_0\Omega_{s0}}\frac{-k_{\zeta
    }^2+k_{\psi}^2{B_0^2r_0^4}z}{4z^{5/4}\sqrt{k_{\zeta
    }^2+k_{\psi}^2{B_0^2r_0^4}z}}6\sin\xi\cos\xi,
\end{equation*}
and
\begin{equation*}
    b_s=\Big(\frac{k_\perp v_{ts}}{\Omega_s}\Big)^2=\frac{v_{ts}^2}{r_0^2\Omega_{s0}^2}\Big[k_{\zeta
    }^2+k_{\psi}^2{B_0^2r_0^4}z\Big]\frac{\cos^6\xi}{z},
\end{equation*}
with $z=1+3\sin^2\xi$.

The $\omega_{*s}^T$ is the same as in the ring dipole case. The
gradient drift ${\bm v}_g$, curvature drift ${\bm v}_c$ and total
drift ${\bm v}_d={\bm v}_g+{\bm v}_c$ can be calculated as
\begin{equation*}
  {\bm v}_g=\frac{1}{m\Omega_s}\mu {\bm b}\times\nabla B=-\frac{v_\perp^2}{2\Omega_s}(\partial_\psi B){\bm
  e}_\zeta,
\end{equation*}
\begin{equation*}
  {\bm v}_c=\frac{1}{\Omega_s}v_\parallel^2\nabla\times {\bm b}=-\frac{v_\parallel^2}{\Omega_s}(\partial_\psi B){\bm
  e}_\zeta,
\end{equation*}
\begin{equation*}
  {\bm v}_d=-\frac{3(\cos^2\theta+1)\sin^5\theta}{r_0\Omega_{s0}(1+3\cos^2\theta)^{2}}(v_\parallel^2+\frac{1}{2}v_\perp^2){{\hat
  e}_\phi},
\end{equation*}
where we have used $\partial_\psi
B=\frac{3(\cos^2\theta+1)}{r^2(1+3\cos^2\theta)^{3/2}}$ and ${\bm
e}_\zeta=\partial_\zeta{\bm r}=r\sin\theta{{\hat e}_\phi}$. And thus
\begin{equation}
  \omega_{Ds}={\bf k_\perp}\cdot{\bm v}_d
  =-\omega_{d0}\frac{3(1-\sin^4\xi)}{(1+3\sin^2\xi)^{2}}\frac{(1+y)}{2}v^2,
\end{equation}
with $\omega_{d0}=\frac{k_\zeta}{r_0^2\Omega_{s0}}$.

We use the same normalization as in the ring dipole case and the
expression for all other variables are the same, except
$\epsilon_n\equiv-\frac{1}{B_0r_0^2}\Big(\frac{\partial \ln
n_0}{\partial \psi}\Big)^{-1}=L_n/B_0r_0^2$ and $k_\psi\to k_\psi
B_0r_0^2$. And hence
\begin{equation*}
    k_\perp^2=(k_{\zeta
    }^2+k_{\psi}^2z)/\cos^6\xi,
\end{equation*}
\begin{equation*}
  k_\perp\rho_{s} = \frac{v}{v_{ts}}\sqrt{\lambda}\sqrt{(k_{\zeta
    }^2+k_{\psi}^2z)}z^{-1/4}\rho_{ts},
\end{equation*}
\begin{equation*}
  \partial_\xi( k_\perp\rho_s) = \frac{v}{v_{ts}}\sqrt{\lambda}\frac{(-k_{\zeta
    }^2+k_{\psi}^2z)}{2z^{5/4}\sqrt{k_{\zeta
    }^2+k_{\psi}^2z}}3\sin\xi\cos\xi\rho_{ts}.
\end{equation*}
\begin{equation*}
    b_s=\rho_{ts}^2\Big[k_{\zeta
    }^2+k_{\psi}^2z\Big]\frac{\cos^6\xi}{z}.
\end{equation*} Normalized curvature
drift frequency is
\begin{equation*}
  \omega_{Ds}
  =-\omega_{ds0}\frac{3(1-\sin^4\xi)}{(1+3\sin^2\xi)^{2}}\frac{(1+y)}{2}\frac{v^2}{v_{ts}^2},
\end{equation*}
where $\omega_{di0}=k_\zeta\omega_0\rho_{ti}\to k_\zeta\rho_{ti}$
and $\omega_{de0}=\tau\frac{q_i}{q_e}\omega_{di0}$.

\section{Bounce average}\label{sec:apd_bounce_avg}

\begin{figure}
 \centering
  \includegraphics[width=8cm]{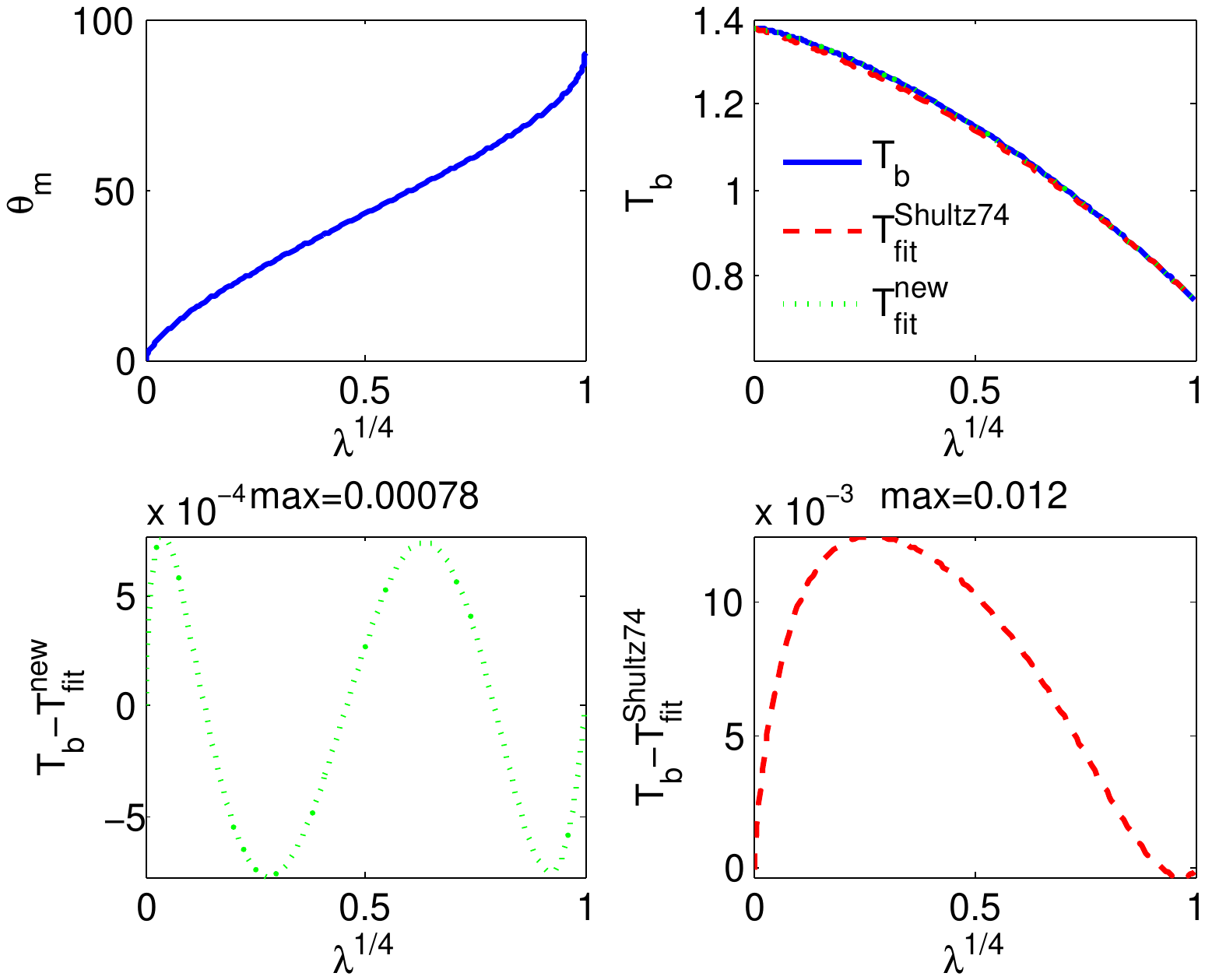}\\
  \caption{Turning point and bounce time.}\label{fig:bounce_integral}
\end{figure}

\begin{figure}
 \centering
  \includegraphics[width=8cm]{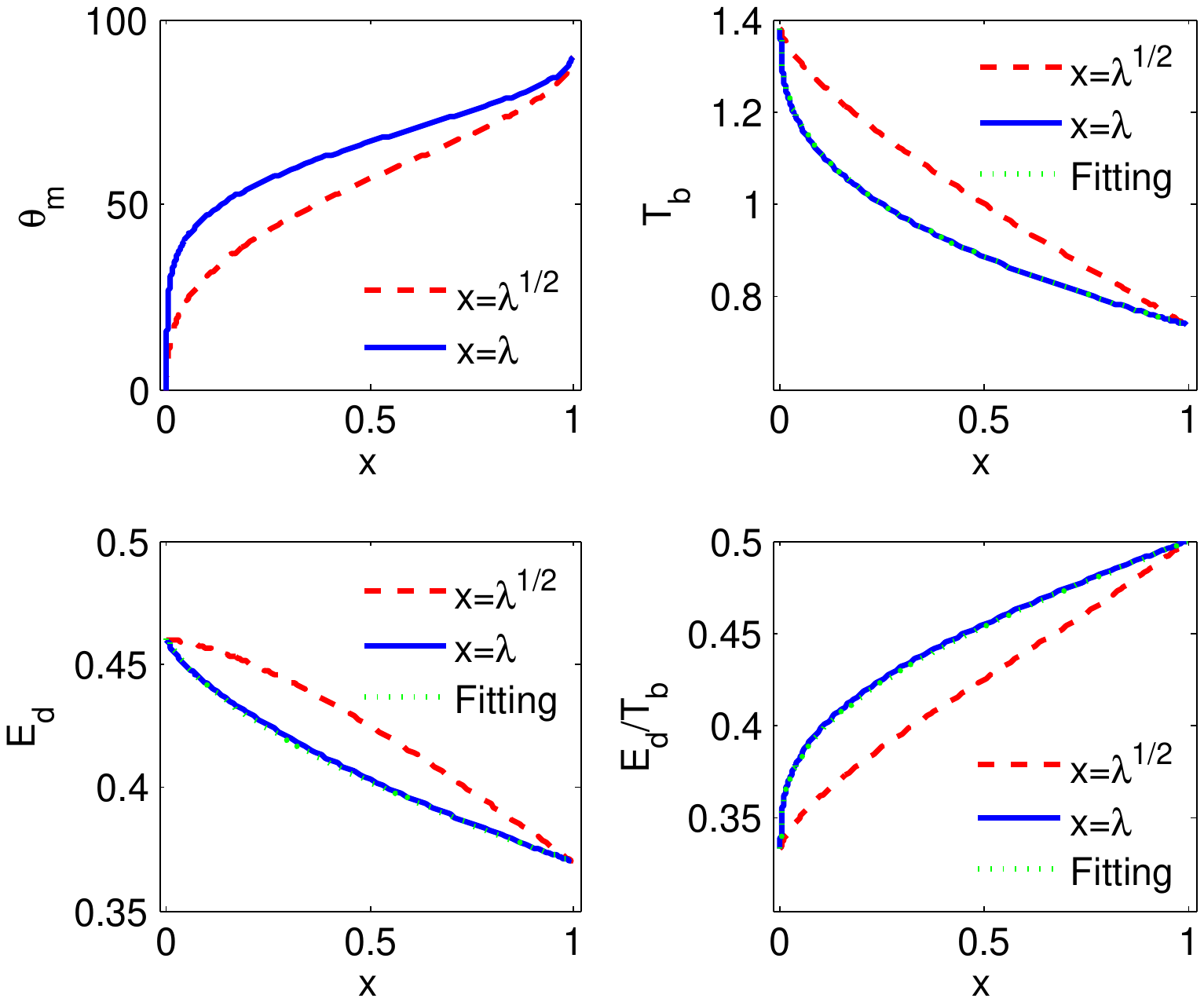}\\
  \caption{Drift integral.}\label{fig:drift_integral}
\end{figure}

Using $E$ and $\mu$ conserved, $v_\parallel=\sqrt{(2/m)(E-\mu
B)}=v\sqrt{1-\lambda B/B_0}$, $v=\sqrt{2E/m}$, the bounce period is
\begin{equation}
  T_m(r_0,E,\lambda)=4\int_{\theta_m}^{\pi/2}\frac{dl}{d\theta}\frac{d\theta}{v_\parallel(\theta)}=\frac{4r_0}{v}T_b(\lambda),
\end{equation}
which gives
\begin{equation}\label{eq:Tb}
  T_b(\lambda)=\int_{\theta_m(\lambda)}^{\pi/2}\frac{\sin\theta(1+3\cos^2\theta)^{1/2}}{\Big[1-
  \lambda\frac{(1+3\cos^2\theta)^{1/2}}{\sin^6\theta}\Big]^{1/2}}d\theta,
\end{equation}
where the turning point $\theta_m(\lambda)$ is determined by
$v_\parallel=0$, i.e,
\begin{equation}
  \frac{\sin^6\theta_m}{(1+3\cos^2\theta_m)^{1/2}}=\lambda.
\end{equation}
For $\lambda=0$,
$T_b(0)=1+(\sqrt{3}/6)\ln(2+\sqrt{3})\simeq1.38017$. For
$\lambda=1$, the period can be calculated by small-amplitude
oscillation $T_b(1)=(\pi/6)\sqrt{2}\simeq0.74048$. Mirror force
$F=\mu\nabla B$, gives $\frac{d^2s}{dt^2}=-\frac{\mu\nabla
B}{m}=-\frac{\mu}{m}(\partial_\chi B) |{\bm e}^{\chi}|$. For deeply
trapped particle, $\lambda\sim1$, $\theta\sim\pi/2$, $ds\sim
r_0d\theta$, and thus
$\frac{d^2s}{dt^2}\simeq-\frac{v^2}{2}\frac{3\cos\theta(5\cos^2\theta+3)}{r(1+3\cos^2\theta)^{3/2}}
\simeq-\frac{v^2}{2}\frac{9s}{r_0^2}$, which gives
$T_b(\lambda=1)=(2\pi/\sqrt{\frac{9v^2}{2r_0^2}})/(4r_0/v)=\frac{\pi\sqrt{2}}{6}$.
A fitting of $T_b(\lambda)$ can be found at Ref.\cite{Schulz1974}
\begin{eqnarray}\nonumber
  T_b(\lambda)&\simeq&
  T_b(0)-\frac{1}{2}[T_b(0)-T_b(1)](\lambda^{1/2}+\lambda^{1/4})\\
  &\simeq&1.3802-0.3198(\lambda^{1/2}+\lambda^{1/4}).
\end{eqnarray}
which gives a max error around $0.012$. We find a better fitting can
be
\begin{equation}
  T_b(\lambda)\simeq
  T_b(0)-\frac{1}{a+b+c}[T_b(0)-T_b(1)](a\lambda^{1/2}+b\lambda^{1/4}+c\lambda^{3/8}),
\end{equation}
with $a=0.380$, $b=0.335$ and $c=1.0$, which gives a max error
around $0.0008$, ten times better than the previous one, see
Fig.\ref{fig:bounce_integral}. The bounce frequency is given by
\begin{equation}
  \omega_b=\frac{2\pi}{T_m}=\frac{\pi
  v}{2r_0}\frac{1}{T_b(\lambda)}.
\end{equation}
In one bounce period, the angular displacement is
\begin{equation}
  \Delta \phi=4\int_{\theta_m}^{\pi/2}\frac{dl}{d\theta}\frac{v_d(\theta)d\theta}{v_\parallel(\theta)
  r(\theta)\sin\theta},
\end{equation}
and thus the angular drift velocity ($B_0=M/r_0^3$, $\Omega_c=eB/m$)
\begin{equation}
  \omega_d=\frac{\Delta \phi}{T_m}=\frac{3mv^2}{eB_0r_0^2}\frac{E_d(\lambda)}{T_b(\lambda)},
\end{equation}
where
\begin{equation}\label{eq:Ed}
  E_d(\lambda)=\int_{\theta_m}^{\pi/2}\frac{\sin^3\theta(1+\cos^2\theta)\Big[1-
  \frac{1}{2}\lambda\frac{(1+3\cos^2\theta)^{1/2}}{\sin^6\theta}\Big]}{(1+3\cos^2\theta)^{3/2}\Big[1-
  \lambda\frac{(1+3\cos^2\theta)^{1/2}}{\sin^6\theta}\Big]^{1/2}}d\theta,
\end{equation}
which agrees with Ref.\cite{Hamlin1961}. Result is shown in
Fig.\ref{fig:drift_integral}. We can also calculate analytically
$E_d(\lambda=0)=[6+\sqrt{3}\ln(2+\sqrt{3})]/18\simeq0.460058$ and
$E_d(\lambda=1)=\pi\sqrt{2}/12$. Compare Eqs.(\ref{eq:Tb}) and
(\ref{eq:Ed}),
$\frac{E_d(\lambda=1)}{T_b(\lambda=1)}=1-\frac{\lambda}{2}=\frac{1}{2}$.

We try the below fitting expression (see also Ref,\cite{Mauel2015}
for similar fitting){\small
\begin{equation}
  T_b(\lambda)\simeq
  T_b(0)+[T_b(1)-T_b(0)]\lambda^{3/8},
\end{equation}
\begin{equation}
  E_d(\lambda)\simeq
  E_d(0)+[E_d(1)-E_d(0)](\lambda^{3/8}+c\lambda^{1/2})/(1+c),~c=-2,
\end{equation}}
which seems can have max error less than $0.002$.
Ref.\cite{Hamlin1961} also gives another two approximations:
$T_b(\lambda)\simeq1.30-0.56\lambda^{1/2}$ and
$E_d(\lambda)/T_b(\lambda)\simeq0.35+0.15\lambda^{1/2}$. In
numerical aspects, we can also try high order polynomial fitting,
e.g., using $\lambda^{1/8}$ as base.

The gyrokinetic dispersion relation in dipole field involve FLR
effect and bounce average, i.e., Bessel function  $\langle
J_0(k_\perp\rho)\rangle_b$, where
$\rho=\frac{mv_\perp}{qB}=\frac{m}{q}\sqrt{\frac{2\lambda
E}{B_0B(\theta)}}=\frac{mv}{q}\sqrt{\frac{\lambda }{B_0B(\theta)}}$,
i.e.,{\small
\begin{eqnarray}\label{eq:J0b}\nonumber
  &&J_{0b}(v,\lambda)=\frac{\int_{\theta_m}^{\pi/2}\frac{dl}{d\theta}\frac{J_0(k_\perp\rho)d\theta}{v_\parallel(\theta)}}
  {\int_{\theta_m}^{\pi/2}\frac{dl}{d\theta}\frac{d\theta}{v_\parallel(\theta)}}
  =\frac{\int_{\theta_m}^{\pi/2}\frac{dl}{d\theta}\frac{J_0\Big(k_\perp\frac{mv}{q}\sqrt{\frac{\lambda }{B_0B(\theta)}}\Big)d\theta}{v_\parallel(\theta)}}
  {\frac{r_0}{v}T_b(\lambda)}\\
  &&=\frac{1}{T_b(\lambda)}\int_{\theta_m(\lambda)}^{\pi/2}\frac{J_0\Big(k_\perp\frac{mv}{q}\sqrt{\frac{\lambda }{B_0B(\theta)}}\Big)\sin\theta(1+3\cos^2\theta)^{1/2}}{\Big[1-\lambda\frac{(1+3\cos^2\theta)^{1/2}}{\sin^6\theta}\Big]^{1/2}}d\theta,
\end{eqnarray}}
Note also that $J_{0b}\equiv\langle J_0(k_\perp\rho)\rangle_b\neq
J_0(\langle k_\perp\rho\rangle_b)$.

{\color{blue}
\begin{figure}
 \centering
  \includegraphics[width=8.0cm]{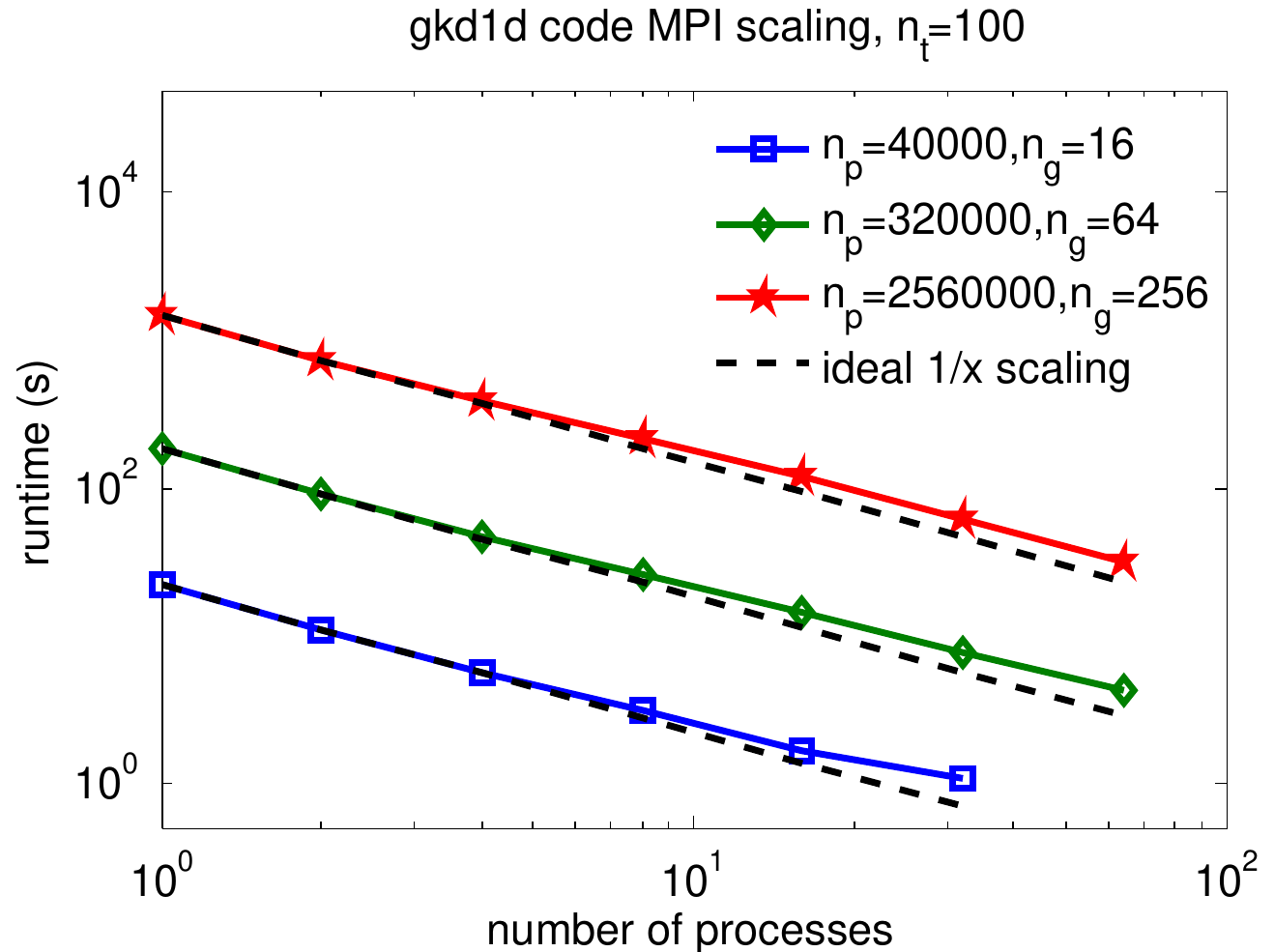}\\
  \caption{Gkd1d Fortran 90 code MPI scaling.
  }\label{fig:gkd1d_mpi_scaling}
\end{figure}

\begin{figure}
 \centering
  \includegraphics[width=8.5cm]{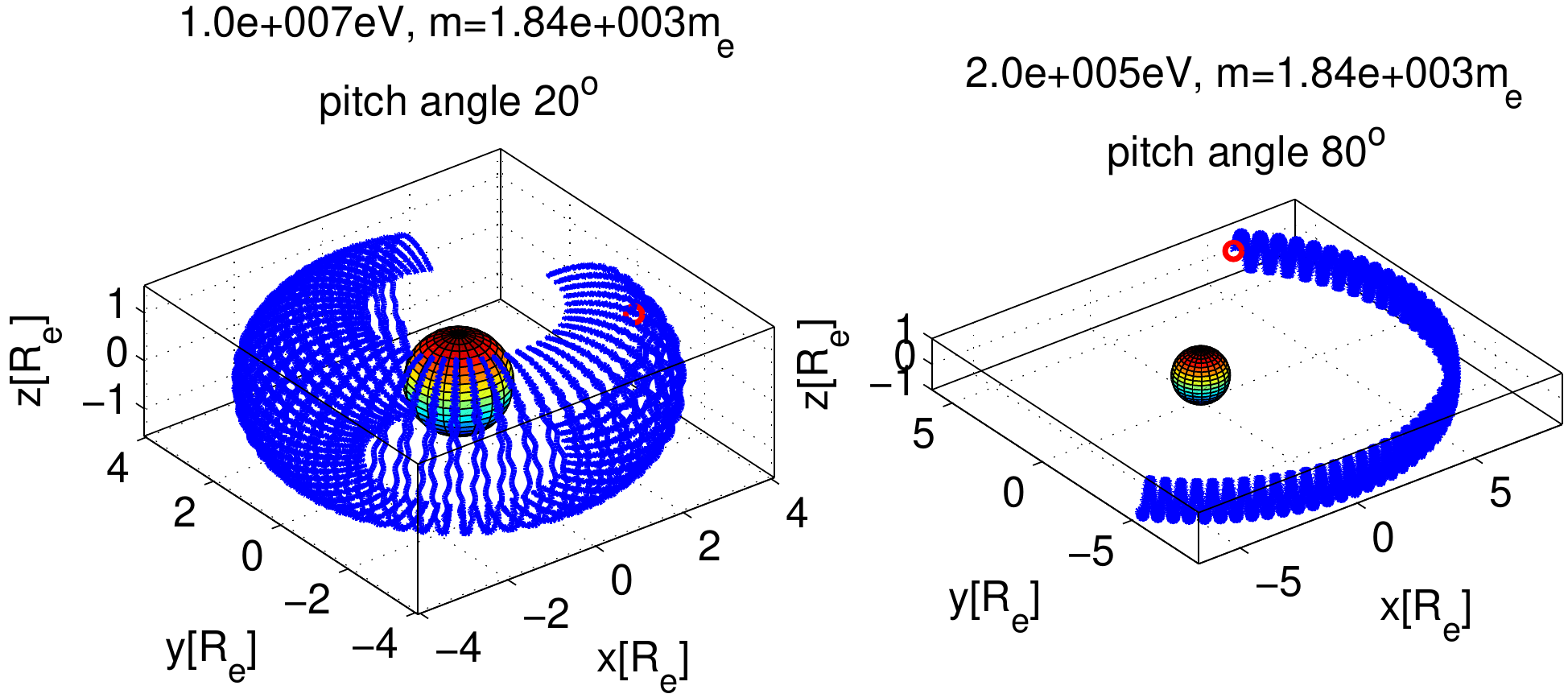}\\
  \caption{Typical charged particle trajectories under ideal dipole field in Earth
  shows good confinement even when the ion energy is large to 10MeV.
  Stochastic motion of particles
due to the collision and turbulence can break this ideal
confinement.
  }\label{fig:orbit_dipole}
\end{figure}

\section{MPI parallelization scaling}
The gkd1d code is written using Fortran90 and with MPI (message
passing interface) parallelization for particles. Thus this code
could handle very large particle numbers. The parallelization
performance is shown in Fig.\ref{fig:gkd1d_mpi_scaling}. In practise
convergence test, we have used more than $n_p=10^7$ particles, and
which is far adequate for most of our simulations. Usually,
$n_p=10^5$ is enough for two species, $s=i,e$; and can even
$n_p\leq10^4$ for adiabatic electron case, i.e., $s=i$.

\section{Typical dipole orbits}
Fig.\ref{fig:orbit_dipole} shows two typical charged particle
trajectories under ideal dipole field in Earth with the magnetic
field $B(x=R_e,y=0,z=0)=3.07\times10^5T$, which shows good
confinement even when the ion energy is large to 10MeV, where $R_e$
is the Earth radius.

\vspace{20pt} Note: The blue text in only in this arXiv version, not
included in the published Phys. Plasmas (2017) version.

}

\end{document}